\documentclass[11pt]{article}
\usepackage{axodraw}
\usepackage{epsfig}
\usepackage{amsfonts}
\usepackage{amsmath}
\usepackage{bbm}
\usepackage{cite}
\allowdisplaybreaks[1]

\input pix.sty

\newcommand{\nS}{n_\rmii{S}}
\newcommand{\nW}{n_\rmii{W}}
\newcommand{\nG}{n_\rmii{G}}

\newcommand{\mZ}{m_\rmii{$Z$}}

\newcommand{\mH}{m_\rmii{$H$}}

\newcommand{\aL}{a^{ }_\rmii{L}}
\newcommand{\aR}{a^{ }_\rmii{R}}
\renewcommand{\eq}{eq.~}
\renewcommand{\eqs}{eqs.~}

\renewcommand{\se}{sec.~}
\renewcommand{\ses}{secs.~}
\renewcommand{\fig}{fig.~}

\newcommand{\tinymsbar}{{\overline{\mbox{\tiny\rm{MS}}}}}

\newcommand{\mD}{m_\rmii{D}}

\newcommand{\alphas}{\alpha_{\rm s}}
\newcommand{\Nf}{N_{\rm f}}
\newcommand{\Nc}{N_{\rm c}}

\newcommand{\gammaE}{\gamma_\rmii{E}}
\newcommand{\rmO}{{\mathcal{O}}}
\newcommand{\bmu}{\bar\mu}

\newcommand{\CF}{C_\rmii{F}}

\newcommand{\aEms}[1]{\alpha_\rmi{E#1}^\tinymsbar}

\def\lsi{\raise0.3ex\hbox{$<$\kern-0.75em\raise-1.1ex\hbox{$\sim$}}}
\def\gsi{\raise0.3ex\hbox{$>$\kern-0.75em\raise-1.1ex\hbox{$\sim$}}}
\newcommand{\lsim}{\mathop{\lsi}}
\newcommand{\gsim}{\mathop{\gsi}}

\newcommand{\nF}{n_\rmii{F}}
\newcommand{\nB}{n_\rmii{B}}
 \renewcommand{\nF}[1]{n_\rmii{F{#1}}}
 \renewcommand{\nB}[1]{n_\rmii{B{#1}}}

\newcommand{\rmii}[1]{{\mbox{\tiny\rm{#1}}}}

\newcommand{\re}{\mathop{\mbox{Re}}}
\newcommand{\im}{\mathop{\mbox{Im}}}

\newcommand{\Tint}[1]{{\hbox{$\sum$}\!\!\!\!\!\!\!\int\,}_{\!\!\!\!\raise-0.9ex\hbox{$\scriptstyle{#1}$}}}
\newcommand{\Tinti}[1]{{{\Sigma}\!\!\!\!\raise0.3ex\hbox{$\int$}_\rmii{${#1}$}}}

\newcommand{\bi}{\begin{itemize}}
\newcommand{\ei}{\end{itemize}}
\newcommand{\hide}[1]{ }

\newcommand{\msl}[1]{\,\slash\!\!\!{#1}\,}
  
\newcommand{\deltabar}{\delta\!\!\!\raise0.7ex\hbox{--}\,}
\def\TAsc(#1,#2)(#3,#4,#5)%
{\SetWidth{2.0}\CArc(#1,#2)(#3,#4,#5)\SetWidth{1.0}}
\def\Lwidth{3}

\def\TAgl(#1,#2)(#3,#4,#5){\SetWidth{2.0}\PhotonArc(#1,#2)(#3,#4,#5){\Lwidth}%
{6.283 #3 mul 360 div #4 #5 sub #4 #5 sub mul sqrt mul Tdensity mul}%
\SetWidth{1.0}}
\def\TLgl(#1,#2)(#3,#4){\SetWidth{2.0}\Photon(#1,#2)(#3,#4){\Lwidth}
{#1 #3 sub #1 #3 sub mul #2 #4 sub #2 #4 sub mul add sqrt Tdensity mul}%
\SetWidth{1.0}}

\renewcommand{\picc}[1]{\;\parbox[c]{60pt}{\begin{picture}(60,30)(0,0)
\SetWidth{1.0}\SetScale{1.3} #1 \end{picture}}\; }
\def\Lwidth{1.3}

\newcommand{\pictiny}[1]{\;\parbox[c]{10pt}{\begin{picture}(10,5)(0,0)
\SetWidth{1.0}\SetScale{1.0} #1 \end{picture}}\; }
\renewcommand{\pic}[1]{\;\parbox[c]{30pt}{\begin{picture}(30,30)(0,-3)
\SetWidth{1.0}\SetScale{0.8} #1 \end{picture}}\;}
\renewcommand{\picb}[1]{\;\parbox[c]{45pt}{\begin{picture}(45,30)(0,-3)
\SetWidth{1.0}\SetScale{0.8} #1 \end{picture}}\;}
\def\indexeta{\pictiny{%
 \SetScale{0.7}
 \SetWidth{3.0} 
 \Lsc(0,5)(10,5)%
}}
\def\naiveRe{\picccc{%
 \SetScale{0.7}
 \SetWidth{3.0} 
 \Lsc(0,15)(10,15)%
 \Asc(25,15)(15,180,360)%
 \Lsc(40,15)(50,15)%
 \SetWidth{1.0} 
 \Agl(25,15)(15,0,180)%
 \Text(44,10)[c]{+}%
 \SetWidth{3.0} 
 \Lsc(70,10)(120,10)%
 \SetWidth{1.0} 
 \Agl(95,20)(10,-90,270)
}}
\def\naiveIm{\picccc{%
 \SetScale{0.7}
 \SetWidth{3.0} 
 \Lsc(0,15)(10,15)%
 \Asc(25,15)(15,180,360)%
 \Lsc(40,15)(50,15)%
 \SetWidth{1.0} 
 \Agl(25,15)(15,0,180)%
 \Line(10,0)(40,30)
 \Text(44,10)[c]{$\sim$}%
 \Line(75,0)(75,30)%
 \SetWidth{3.0} 
 \Lsc(80,10)(110,10)%
 \SetWidth{1.0} 
 \Lgl(95,10)(95,30)%
 \Line(115,0)(115,30)%
 \Text(88,18)[c]{\small 2}
}}
\def\htlRe{\picccc{%
 \SetScale{0.7}
 \SetWidth{3.0} 
 \Lsc(0,15)(10,15)%
 \Asc(25,15)(15,180,360)%
 \Lsc(40,15)(50,15)%
 \SetWidth{1.0} 
 \Agl(25,15)(15,0,180)%
 \Text(44,10)[c]{+}%
 \SetWidth{3.0} 
 \Lsc(70,10)(120,10)%
 \SetWidth{1.0} 
 \Agl(95,20)(10,-90,270)%
 \COval(25,30)(4,4)(0){Black}{Black}%
 \COval(95,30)(3,3)(0){Black}{Black}
}}
\def\htlIm{\picccc{%
 \SetScale{0.7}
 \SetWidth{3.0} 
 \Lsc(0,15)(10,15)%
 \Asc(25,15)(15,180,360)%
 \Lsc(40,15)(50,15)%
 \SetWidth{1.0} 
 \Agl(25,15)(15,0,180)%
 \Line(10,0)(40,30)
 \Text(44,10)[c]{$\sim$}%
 \Line(75,0)(75,30)%
 \SetWidth{3.0} 
 \Lsc(80,8)(112,8)%
 \SetWidth{1.0} 
 \Lgl(96,8)(96,28)%
 \Line(117,0)(117,30)%
 \Text(88,18)[c]{\small 2}
 \COval(25,30)(4,4)(0){Black}{Black}%
 \COval(96,19)(3,3)(0){Black}{Black}
 \Line(96,28)(104,30)%
 \Line(96,28)(88,30)%
}}
\def\pair{\picccc{%
 \SetScale{0.7}
 \SetWidth{3.0} 
 \Lsc(0,14)(120,14)%
 \Lsc(0,-8)(120,-8)%
 \SetWidth{1.0} 
 \Agl(60,15)(15,0,180)%
 \Lgl(100,-7)(100,13)%
 \COval(60,30)(4,4)(0){Black}{Black}%
 \COval(100,3)(3,3)(0){Black}{Black}%
 \LongArrow(7,-6)(7,12)%
 \LongArrow(7,12)(7,-6)%
 \Text(0,1.5)[c]{$r$}%
}}
\def\htlDM{\piccccc{%
 \SetScale{0.7}
 \SetWidth{3.0} 
 \Line(0,15)(10,15)%
 \Asc(25,15)(15,180,360)%
 \Line(40,15)(50,15)%
 \SetWidth{1.0} 
 \Aaqu(25,15)(15,5,90)%
 \Aaqu(25,15)(15,90,175)%
 \Line(10,0)(40,30)
 \COval(25,30)(4,4)(0){Black}{Black}%
 \Text(44,10)[c]{$\sim$}%
 \Line(75,0)(75,30)%
 \SetWidth{3.0} 
 \Lsc(98,8)(112,8)%
 \Line(80,8)(96,8)%
 \SetWidth{1.0} 
 \Line(96,9.5)(96,28)%
 \Line(117,0)(117,30)%
 \Text(87,18)[c]{\small 2}
 \COval(96,19)(3,3)(0){Black}{Black}%
 \Lqu(96,28)(108,30)%
 \Lgl(96,28)(84,30)%
 \Text(94,10)[c]{$+$}%
 \Line(145,0)(145,30)%
 \SetWidth{3.0} 
 \Lsc(168,8)(182,8)%
 \Line(150,8)(166,8)%
 \SetWidth{1.0} 
 \Line(166,9.5)(166,28)%
 \Line(187,0)(187,30)%
 \Text(136,18)[c]{\small 2}
 \COval(166,19)(3,3)(0){Black}{Black}%
 \Lgl(166,28)(178,30)%
 \Lqu(166,28)(154,30)%
}}
\def\mprocA{\pic{%
 \SetWidth{3.0}
 \Lsc(0,30)(15,15)%
 \Lsc(0,0)(15,15)%
 \SetWidth{1.0}
 \Lgl(15,15)(30,30)%
 \Lgl(15,15)(30,0)%
}}
\def\mprocB{\pic{%
 \SetWidth{3.0}
 \Lsc(0,0)(15,5)%
 \Lsc(0,30)(15,25)%
 \Lsc(15,5)(15,25)%
 \SetWidth{1.0}
 \Lgl(15,4)(30,0)%
 \Lgl(15,25)(30,30)%
}}
\def\mprocC{\picb{%
 \SetWidth{3.0}
 \Lsc(0,0)(15,5)%
 \Lsc(0,30)(15,25)%
 \Lsc(15,5)(15,25)%
 \SetWidth{1.0}
 \Lgl(15,4)(25,12.6)%
 \Lgl(30,17)(45,30)%
 \Lgl(15,25)(45,0)%
}}
\def\mprocD{\picb{%
 \SetWidth{3.0}
 \Lsc(0,30)(15,15)%
 \Lsc(0,0)(15,15)%
 \SetWidth{1.0}
 \Lgl(15,15)(30,15)%
 \Lgl(30,15)(45,30)%
 \Lgl(45,0)(30,15)%
}}
\def\mprocE{\pic{%
 \SetWidth{3.0}
 \Lsc(0,30)(15,15)%
 \Lsc(0,0)(15,15)%
 \SetWidth{1.0}
 \Lsc(15,15)(30,30)%
 \Lsc(15,15)(30,0)%
}}
\def\mprocG{\pic{%
 \SetWidth{3.0}
 \Lsc(0,0)(15,5)%
 \Lsc(0,30)(15,25)%
 \Line(15,4)(15,25)%
 \SetWidth{1.0}
 \Lqu(15,4)(30,0)%
 \Laqu(15,25)(30,30)%
}}
\def\mprocH{\picb{%
 \SetWidth{3.0}
 \Lsc(0,0)(15,5)%
 \Lsc(0,30)(15,25)%
 \Line(15,5)(15,25)%
 \SetWidth{1.0}
 \Line(15,4)(25,12.6)%
 \Laqu(30,17)(45,30)%
 \Line(15,25)(22.5,18.75)%
 \Lqu(22.5,18.75)(45,0)%
}}
\def\mprocI{\picb{%
 \SetWidth{3.0}
 \Lsc(0,30)(15,15)%
 \Lsc(0,0)(15,15)%
 \SetWidth{1.0}
 \Lgl(15,15)(30,15)%
 \Laqu(30,15)(45,30)%
 \Lqu(30,15)(45,0)%
}}
\def\mprocJ{\pic{%
 \SetWidth{3.0}
 \Lsc(0,0)(15,5)%
 \Lsc(0,30)(15,25)%
 \Line(15,4)(15,25)%
 \SetWidth{1.0}
 \Laqu(15,4)(30,0)%
 \Laqu(15,25)(30,30)%
}}
\def\mprocK{\picb{%
 \SetWidth{3.0}
 \Lsc(0,0)(15,5)%
 \Lsc(0,30)(15,25)%
 \Line(15,5)(15,25)%
 \SetWidth{1.0}
 \Line(15,4)(25,12.6)%
 \Laqu(30,17)(45,30)%
 \Line(15,25)(22.5,18.75)%
 \Laqu(22.5,18.75)(45,0)%
}}
\def\mprocL{\pic{%
 \SetWidth{3.0}
 \Lsc(0,0)(15,5)%
 \Lsc(0,30)(15,25)%
 \Line(15,4)(15,25)%
 \SetWidth{1.0}
 \Lqu(15,4)(30,0)%
 \Lqu(15,25)(30,30)%
}}
\def\mprocM{\picb{%
 \SetWidth{3.0}
 \Lsc(0,0)(15,5)%
 \Lsc(0,30)(15,25)%
 \Line(15,5)(15,25)%
 \SetWidth{1.0}
 \Line(15,4)(25,12.6)%
 \Lqu(30,17)(45,30)%
 \Line(15,25)(22.5,18.75)%
 \Lqu(22.5,18.75)(45,0)%
}}
\def\mprocN{\pic{%
 \SetWidth{3.0}
 \Line(0,0)(15,5)%
 \Lsc(0,30)(15,25)%
 \Lsc(15,5)(15,25)%
 \SetWidth{1.0}
 \Lqu(15,4)(30,0)%
 \Lgl(15,25)(30,30)%
}}
\def\mprocO{\picb{%
 \SetWidth{3.0}
 \Lsc(0,30)(15,15)%
 \Line(0,0)(15,15)%
 \SetWidth{1.0}
 \Lqu(15,15)(30,15)%
 \Lgl(30,15)(45,30)%
 \Lqu(30,15)(45,0)%
}}
\def\mprocP{\pic{%
 \SetWidth{3.0}
 \Line(0,0)(15,5)%
 \Lsc(0,30)(15,25)%
 \Lsc(15,5)(15,25)%
 \SetWidth{1.0}
 \Laqu(15,4)(30,0)%
 \Lgl(15,25)(30,30)%
}}
\def\mprocQ{\picb{%
 \SetWidth{3.0}
 \Lsc(0,30)(15,15)%
 \Line(0,0)(15,15)%
 \SetWidth{1.0}
 \Laqu(15,15)(30,15)%
 \Lgl(30,15)(45,30)%
 \Laqu(30,15)(45,0)%
}}
\def\mprocR{\picb{%
 \SetWidth{3.0}
 \Lsc(0,30)(15,15)%
 \Line(0,0)(15,15)%
 \SetWidth{1.0}
 \Lqu(15,15)(30,15)%
 \Lsc(30,15)(45,30)%
 \Lqu(30,15)(45,0)%
}}
\def\mprocS{\picb{%
 \SetWidth{3.0}
 \Lsc(0,30)(15,15)%
 \Line(0,0)(15,15)%
 \SetWidth{1.0}
 \Laqu(15,15)(30,15)%
 \Lsc(30,15)(45,30)%
 \Laqu(30,15)(45,0)%
}}
\def\mprocT{\pic{%
 \SetWidth{3.0}
 \Line(0,0)(15,5)%
 \Line(0,30)(15,25)%
 \Lsc(15,5)(15,25)%
 \SetWidth{1.0}
 \Lqu(15,4)(30,0)%
 \Laqu(15,25)(30,30)%
}}
\def\mprocU{\picb{%
 \SetWidth{3.0}
 \Line(0,0)(15,5)%
 \Line(0,30)(15,25)%
 \Lsc(15,5)(15,25)%
 \SetWidth{1.0}
 \Line(15,4)(25,12.6)%
 \Laqu(30,17)(45,30)%
 \Line(15,25)(22.5,18.75)%
 \Lqu(22.5,18.75)(45,0)%
}}
\makeatletter \@addtoreset{equation}{section} \makeatother
\renewcommand{\theequation}{\arabic{section}.\arabic{equation}}
\makeatletter
\renewcommand\section{\@startsection {section}{1}{\z@}%
                                   {-5.5ex \@plus -1ex \@minus -.2ex}
                                   {2.3ex \@plus.2ex}%
                                   {\normalfont\large\bfseries}}
\renewcommand\subsection{\@startsection{subsection}{2}{\z@}%
                                     {-3.25ex\@plus -1ex \@minus -.2ex}%
                                     {1.5ex \@plus .2ex}%
                                     {\normalfont\normalsize\bfseries}}
\renewcommand\thesection {\@arabic\c@section}
\renewcommand\thesubsection   {\thesection.\@arabic\c@subsection}
\renewcommand{\@seccntformat}[1]{%
\csname the#1\endcsname.\hspace{1.0em}}
\makeatother

\begin{document}

\flushbottom

\begin{titlepage}

\begin{flushright}
April 2018
\vspace*{1cm}
\end{flushright} 
\begin{centering}

\vfill

{\Large{\bf
 Thermal dark matter co-annihilating \\[3mm]
 with a strongly interacting scalar 
}} 

\vspace{0.8cm}

S.~Biondini and 
M.~Laine 

\vspace{0.8cm}

{\em
AEC, Institute for Theoretical Physics, 
University of Bern, \\ 
Sidlerstrasse 5, CH-3012 Bern, Switzerland\\} 

\vspace*{0.8cm}

\mbox{\bf Abstract}

\end{centering}

\vspace*{0.3cm}
 
\noindent
Recently many investigations have considered Majorana dark matter
co-annihilating with bound states formed by a strongly interacting
scalar field.  However only the gluon radiation contribution to bound
state formation and dissociation, which at high temperatures is
subleading to soft $2\to 2$ scatterings, has been included.  Making
use of a non-relativistic effective theory framework and solving a
plasma-modified Schr\"odinger equation, we address the effect of soft
$2\to 2$ scatterings as well as the thermal dissociation of bound
states. We argue that the mass splitting between the Majorana and
scalar field has in general both a lower and an upper bound, and that
the dark matter mass scale can be pushed at least up to $5...6$~TeV.

\vfill

 
\noindent

\vfill

\end{titlepage}

%
\section{Introduction}

Negative results from 
direct and indirect detection experiments and collider searches 
pose a challenge for many minimal 
dark matter models. This has led to the construction 
of less minimal models. In the latter, the cross
sections probed experimentally are not directly
related to the cross section affecting freeze-out dynamics 
in the early universe. 
Therefore experimental bounds might be respected 
while at the same time
maintaining the correct cosmological abundance. 

If the dark matter particles are massive and interact strongly
enough with the Standard Model to have been in equilibrium with it
at some time in the early universe, 
the basic feature that is needed for the above task is  
a strongly temperature-dependent annihilation cross section. 
At low temperatures, the cross section needs to be very small, 
to satisfy the non-observation bounds from indirect detection. In the 
early universe, the cross section needs to be large enough to 
keep dark matter in chemical equilibrium for a long while, 
reducing its number density and thereby evading overclosure
of the universe. 

An example of a possible scenario along these lines is  to postulate a model
in which the dark sector consists of two particle species. 
The lighter one is the true dark matter, long-lived and interacting
very weakly. In contrast, the heavier one could interact strongly
and act as an efficient dilution channel for the overall abundance
at high temperatures (cf.,\ e.g.,\ ref.~\cite{eg}). 

If the heavy species interacts strongly, as in QCD, this scenario
could lead to rather rich phenomenology. Strongly interacting particles
form generally bound states. Because of the associated binding energy, 
their thermal abundance is larger than that for the same particles
in a scattering state. Bound states may annihilate efficiently
because the two particles are close to each other.  
Though often alluded to previously, a more intensive
study of bound-state effects on the freeze-out dynamics has 
only started a few years ago 
(cf.,\ e.g.,\ refs.~\cite{old32,old4}).

Recently, we have participated in developing 
a non-perturbative formalism for addressing 
the thermal annihilation of non-relativistic
particles~\cite{4quark_lattice,threshold}. 
The formalism was already applied to 
a first full model, which did not include strongly interacting
particles but nevertheless displayed weakly
bound states~\cite{idm}. The purpose of the 
current paper is to apply the formalism to a strongly interacting 
model that has been much discussed in recent literature.

Our plan is as follows. 
Having introduced the model in \se\ref{se:model}, 
we review some salient
features concerning its thermal behaviour
in \se\ref{se:thermal}. The main technical ingredients 
of our analysis are specified in \se\ref{se:ingredients}:
the operators responsible for the hard annihilation event; 
the spectral functions describing the soft initial-state effects
that influence the annihilation; as well as generalized
``Sommerfeld factors'' which capture the effect both of bound and
scattering states on the thermal annihilation cross sections. 
The cosmological evolution equations are integrated 
in \se\ref{se:numerics}, whereas 
conclusions and an outlook are offered in \se\ref{se:concl}.

%
\section{Model}
\la{se:model}

The model considered consists of the Standard Model extended by 
a gauge singlet Majorana fermion ($\chi$) as well as a scalar field
($\eta$) which 
is singlet in SU$^{ }_\rmii{L}$(2) but carries non-trivial QCD 
and hypercharge quantum numbers.\footnote{%
 An SU$^{ }_\rmii{L}$(2) doublet $\eta$ would lead to 
 similar results but a somewhat more complicated analysis~\cite{giv}.
 } 
The Majorana fermion is chosen
as the dark matter particle, given that its low-energy scattering
cross section is naturally suppressed, 
being $p$-wave at tree level~\cite{hg}.  
In the MSSM language, 
the Majorana fermion could be a bino-like neutralino and the scalar 
a right-handed stop or sbottom. However, for generality 
we do not fix couplings to their MSSM values.  
The hypercharge coupling of the scalar is generally omitted, 
as its effects are subleading compared with QCD effects.  

The Lagrangian for this extension of the Standard Model can be 
expressed as 
\ba
 \mathcal{L} & = & 
 \mathcal{L}^{ }_\rmii{SM} + 
 \fr12\, \bar\chi \bigl( i \msl{\partial} - M \bigr) \chi 
 + (D^{ }_\mu \eta)^\dagger D^\mu \eta 
 - M_\eta^2\, \eta^\dagger \eta 
 - \lambda^{ }_2 (\eta^\dagger \eta)^2 
 \nn 
 & - & \lambda^{ }_3\, \eta^\dagger \eta\, H^\dagger H 
 - y\,  \eta^\dagger \bar{\chi} \aR q 
 - y^* \bar{q} \aL \chi\, \eta
 \;. \la{L}
\ea
The notation 
$\lambda^{ }_1$
is reserved for the self-coupling of the Higgs doublet
($H$).
The chiral projectors 
$\aL = (\mathbbm{1} - \gamma_{5}^{ })/2$,
$\aR = (\mathbbm{1} + \gamma_{5}^{ })/2$
imply that $\chi$ only interacts with SU$^{ }_\rmii{L}$(2) 
singlet projections of quarks. We assume that 
the Yukawa coupling $y$ couples dominantly to one quark flavour only. 
The Yukawa coupling determining the mass of that flavour
is denoted by~$h$, 
and the strong gauge coupling by $g^{ }_s$. The free parameters
of the model are then the two mass scales\footnote{%
 More precisely, $M$ and $M^{ }_\eta$ refer to the renormalized parts of 
 the masses appearing in the non-relativistic effective theory
 for $\chi$ and $\eta$. The non-perturbative QCD contribution to 
 $M^{ }_\eta$ is of the order $\rmO(\mbox{GeV}/\mbox{TeV}) \sim 10^{-3}$
 which is smaller than the effects that we discuss below. 
 }
$M$ and 
$\Delta M \equiv M^{ }_\eta - M$ as well as the two 
``portal'' couplings $\lambda^{ }_3$ and $|y|^2$ that 
are assumed to be small at the $\msbar$ scale $\bmu \sim 2M$. 

In the MSSM context, the importance of co-annihilations in such
a model was stressed long ago~\cite{eg}. Sommerfeld enhancements from
QCD interactions were included in refs.~\cite{eoz,hhk}, however
without the consideration of bound states. 
Similar theoretical ingredients
were applied to  
the generalized model in ref.~\cite{ipsv}. 
Direct, indirect and collider constraints on 
the generalized model were reviewed in ref.~\cite{giv}. 
More recently, bound-state effects have been approximately 
included in this model~\cite{ll,mrss,klz}, 
as a single additional degree of freedom 
in a set of Boltzmann equations, 
a treatment which we aim to improve upon in the following.

%
\section{Parametric forms of thermal masses and interaction rates}
\la{se:thermal}

The coloured scalars are responsible for
most of the annihilations during thermal freeze-out.
We start by reviewing the thermal mass corrections and interaction
rates that they experience. The important point is that, 
because of Bose enhancement, the gluonic contributions are infrared (IR)
sensitive, and need to be properly resummed for a correct result. 

As a first step, consider a naive (i.e.\ unresummed) computation of 
the self-energy of the coloured scalar. Evaluating the (retarded) 
self-energy at the on-shell point yields (the line ``$\indexeta\!\!$'' stands
for~$\eta$ and the wiggly line for a gluon)
\ba
 \naiveRe \hspace*{-5mm} \Rightarrow
 \frac{\re \Pi^{ }_\rmii{R}}{2 M^{ }_{\eta}} & = &
  \frac{g_s^2 \CF^{ } T^2}{12 M^{ }_\eta}
 \;, \la{naive_re} \\ 
 \naiveIm \hspace*{-5mm} \Rightarrow
 \frac{\im \Pi^{ }_\rmii{R}}{2 M^{ }_{\eta}} & = & 0
 \;, \la{naive_im}
\ea
where $\CF^{ } \equiv (\Nc^2 - 1)/(2\Nc^{ })$.
The real part is analogous to that for a heavy fermion~\cite{dhr}. 
The imaginary part vanishes because there is no phase space for 
the $1\leftrightarrow 2$ process. 

However, at high temperatures 
{\em these naive results are misleading}. Perhaps the 
simplest way to see this is to replace the scalar in the loop by a particle
with a different mass, $M^{ }_\eta + \Delta M$, and consider the case
$\Delta M \ll \pi T \ll M^{ }_\eta$. Then it can be verified that 
$\re\Pi^{ }_\rmii{R}/M^{ }_\eta$ is modified by a correction of order
$\sim g_s^2 \CF^{ }\Delta M$, and $\im\Pi^{ }_\rmii{R}/M^{ }_\eta$
by a correction of order 
$\sim g_s^2 \CF^{ }|\Delta M| \nB{}(|\Delta M|)
 \approx g_s^2 \CF^{ } T 
$,
where $\nB{}$ is the Bose distribution. 
In other words, the result in \eq\nr{naive_im} seems to change qualitatively
because Bose enhancement of the soft contribution compensates
against the phase-space suppression.

The {\em correct} treatment of the Bose-enhanced IR contribution requires 
resummation. The heavy scalars are almost static, and interact 
mostly with colour-electric fields ($A^a_0$). In a plasma, 
colour-electric fields get Debye screened. We denote the Debye mass
by $\mD^{ }$. Parametrically, $\mD^{ }\sim g^{ }_sT$, 
where $g^{ }_s\equiv \sqrt{4\pi\alphas^{ }}$.
The proper inclusion of Debye screening in a gauge
theory requires Hard Thermal Loop (HTL) 
resummation~\cite{ht1,ht2,ht3,ht4}. Recomputing
the 1-loop self-energy with HTL propagators, and setting $\Delta M \to 0$
since IR sensitivity has now been regulated, we get 
(here $p \equiv |\vec{p}|$ and a blob stands for a HTL-resummed propagator)
\ba
 \htlRe \hspace*{-5mm} \Rightarrow 
 \frac{\re \Pi^{*}_\rmii{R}}{2 M^{ }_{\eta}} & = & 
   \frac{g_s^2 \CF^{ } T^2}{12 M^{ }_\eta}
 + \frac{g_s^2 \CF^{ }}{2}
  \int_{\vec{p}} \frac{1}{p^2 + \mD^2} 
 \nn & = & 
   \frac{g_s^2 \CF^{ } T^2}{12 M^{ }_\eta}
 - \frac{g_s^2 \CF^{ } \mD^{ }}{8\pi}
 \;, \la{htl_re} \\ 
 \htlIm \hspace*{-5mm} \Rightarrow
 \frac{\im \Pi^{*}_\rmii{R}}{2 M^{ }_{\eta}} & = & 
 - \frac{g_s^2 \CF^{ }}{2 }
 \int_{\vec{p}} 
 \frac{\pi T \mD^2}{p\,(p^2 + \mD^2)^2}
 \nn  & = &
  - \frac{g_s^2 \CF^{ } T }{8\pi}
 \;. \la{htl_im}
\ea
The new contribution in \eq\nr{htl_re}, 
originating from the Debye-screened Coulomb self-energy, 
is known as the Salpeter correction 
(cf.\ ref.~\cite{lsb} for a review). 
It dominates over the other mass correction if 
$T \lsim g^{ }_s M^{ }_\eta$, which is generally the case.
The imaginary part in \eq\nr{htl_im}, 
i.e.\ the interaction rate, reflects fast colour 
and phase-changing $2\to2$ scatterings
off light medium particles.
It was first derived for the case of a heavy quark~\cite{ht1}. 

We finally replace the coloured scalar by a pair of heavy scalars, 
separated by a distance~$r$. The HTL-resummed computation of the
thermal mass correction (``static potential'') 
and interaction rate as a function of $r$
was carried out in refs.~\cite{imV,bbr,jacopo}. At leading
non-trivial order the result can be expressed as 
\ba
   G(r,t) 
    & \stackrel{t\to+\infty}{\sim} &  
   G(r,0)\, \exp\bigl\{-i [V(r) - i \Gamma(r)] t\bigr\}
 \;,  \la{ex_G} \\[1mm]
 \raise2ex\hbox{\pair} 
 \hspace*{-5mm}
 V(r) & = & - \frac{g_s^2 \CF^{ }}{4\pi}
 \biggl[ 
  \mD^{ } + \frac{\exp(-\mD^{ } r)}{r}
 \biggr]
 \;, \la{ex_V} \\[1mm] 
 \Gamma(r) & = & \frac{g_s^2 \CF^{ } T}{2\pi}
 \int_0^\infty \! \frac{{\rm d} z \, z}{(z^2 +1)^2}
 \biggl[
   1 - \frac{\sin(z \mD^{ }r)}{z \mD^{ }r} 
 \biggr]
 \;. \la{ex_Gamma}
\ea
As a crosscheck, for $r\to \infty$ twice the results of 
\eqs\nr{htl_re} and \nr{htl_im} are reproduced.

The interaction rate in \eq\nr{ex_Gamma} can again be traced back to 
$2\to 2$ scatterings. 
At short distances, up to logarithms, 
$\Gamma(r) \sim g_s^2 \CF^{ } T \mD^2 r^2$.  
This can be compared with the $1\leftrightarrow 2$
gluon radiation contribution,
$\sim g_s^2 \CF^{  } (\Delta E)^3 r^2 \nB{}(\Delta E)$~\cite{jacopo}, 
where $\Delta E$ is the energy difference between the singlet and octet
potentials. At high temperatures, when $\mD^{ },\pi T \gg \Delta E$, 
the $2\to 2$ contribution dominates over the 
$1 \leftrightarrow 2$ one. 

In order to determine the spectral function of the scalar pair, 
characterizing the states that appear in the scalar-antiscalar sector
of the Fock space, $V(r)$ and $\Gamma(r)$ can be inserted into a 
time-dependent Schr\"odinger equation
satisfied by the appropriate Green's function~\cite{resum}. 
More details are given in \se\ref{se:ingredients}. 
We have checked numerically that,  in accordance with theoretical 
expectations~\cite{jacopo2},  
the states originating from this solution respect the qualitative
pattern seen above for $\Gamma(r)$, namely that 
at high temperatures the width from 
$2\to 2$ reactions dominates over the gluo-dissociation 
contribution. 

We close this section by considering another essential ingredient
of the framework, namely the rate at which Majorana dark matter 
particles convert into the coloured scalars. 
Once more, this rate is dominated by $2\to 2$ scatterings, and  
obtaining the correct result requires HTL resummation. Setting
for simplicity the external momentum to zero, 
we find (the thick line is the Majorana fermion and 
the arrowed line the quark flavour with which it interacts, 
treated for simplicity as massless in vacuum which is a good
approximation if $m^{ }_\rmi{vac} \lsim \pi T$)
\ba
 \hspace*{-10mm}
 \htlDM  
 \hspace*{1mm} \Rightarrow \; 
 \im \Sigma^{*}_\rmii{R}
 & = & 
 - \frac{|y|^2 \Nc^{ }}{8 M}
 \int_{\vec{p}} \frac{\pi m_q^2\,
  \nF{}\bigl( \Delta M + \frac{p^2}{2M} \bigr)}
 {p (p^2 + m_q^2)}
 \la{htlDMpre} \\
 &  \approx & 
 - \frac{|y|^2 \Nc^{ } m_q^2}{64\pi M}
 \ln\biggl( \frac{1.76388\, M T}{m_q^2} \biggr)
 \;,   
 \la{htlDM}
\ea
where the last line applies under the assumption
$ \Delta M \ll m^{ }_q, \pi T \ll \sqrt{T M} $. 
The thermal quark mass $m_q^{ }$, originating from the phase space 
integral of the light plasma particles off which the $2\to 2$ 
scattering takes place, is 
\be
 m_q^2 = 2 g_s^2 \CF^{ }
 \int_{\vec{q}} \frac{\nB{}(q)+\nF{}(q)}{q} 
 = \frac{g_s^2 T^2 \CF^{ }}{4}
 \;. \la{mq}
\ee
The rate in \eqs\nr{htlDMpre} and \nr{htlDM} is faster than the Hubble rate
in a broad temperature range, e.g.\ down to 
$M/T\gsim 3000$ for $y=0.3$ and 
$\Delta M / M \lsim 0.01$. It does fall out of equilibrium 
when $T \ll \Delta M$, however transitions to 
virtual bound-state constituents may continue and form 
presumably the relevant concern. 
Non-equilibrium effects have been discussed
in ref.~\cite{ghlv}. 

%
\section{Quantitative framework for estimating the annihilation rate}
\la{se:ingredients}

We now present a framework for computing 
(co-)annihilation rates in the model of \se\ref{se:model}. 

%
\subsection{Non-relativistic fields}

The basic premise of our framework
is to make use of the non-relativistic
approximation, assuming that $\pi T$, $m^{ }_\rmi{top}$, $\Delta M \ll M$, 
where $M$ is the dark matter mass and $\Delta M = M^{ }_\eta - M$
is the mass splitting within the dark sector. 
This simplification opens up the avenue to 
a non-relativistic effective field theory 
investigation of soft initial-state effects. 

In the non-relativistic limit, the interaction picture field operator of 
the coloured scalar is expressed as 
\be
 \eta = \frac{1}{\sqrt{2M^{ }_\eta}}
 \Bigl( \phi\, e^{-i M^{ }_\eta t}
 + \varphi^\dagger\, e^{ i M^{ }_\eta t } \Bigr) 
 \;, 
 \quad
 \eta^\dagger = \frac{1}{\sqrt{2M^{ }_\eta}}
 \Bigl( \varphi\, e^{-i M^{ }_\eta t} 
 + \phi^\dagger\, e^{ i M^{ }_\eta t } \Bigr) 
 \;. \la{eta}
\ee
The non-relativistic fields $\phi$ and $\varphi^\dagger$ transform 
in the fundamental representation of SU($\Nc$), with colour indices
denoted by $\alpha,\beta,\gamma,\delta,...\,$. 
The Majorana spinor
$\chi$ is simplest to handle by choosing the standard representation
for the Dirac matrices, i.e.\ 
$\gamma^0 = \mathop{\mbox{diag}}(\mathbbm{1},-\mathbbm{1})$. Then 
\be
 \chi \; = \; 
 \left( 
  \begin{array}{c}
    \psi\, e^{-i M t} \\ -i \sigma^{ }_2 \psi^*\, e^{i M t}
  \end{array} 
 \right)
 \;, \quad
 \bar{ \chi } \; = \; 
 \left( 
    \psi^\dagger\, e^{i M t} \;,
   - \psi^T i \sigma^{ }_2 \, e^{-i M t}
 \right)
 \;, \la{chi}
\ee
where the Grassmannian spinor $\psi$ has two spin 
components, labelled by $p,q,r,s,...\,$. 
Only the left-chiral projection of $\chi$ participates
in interactions according to \eq\nr{L}. 

In the following, we generally set $M^{ }_\eta \to M$ whenever possible. 
The influence of $\Delta M \neq 0$ (and its thermal modification) 
is discussed in \se\ref{se:splitting}. 

%
\subsection{Imaginary parts of 4-particle operators}
\la{ss:im}

%
\begin{figure}[t]
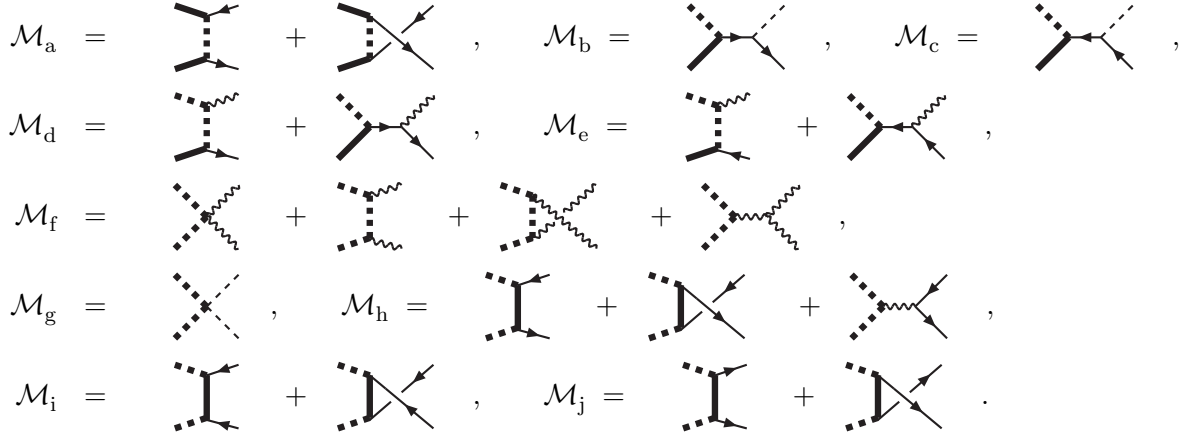


\begin{eqnarray*}
 \mathcal{M}^{ }_\rmi{a} 
 & = & 
 \hspace*{0.45cm}
 \mprocT
 \;\; + \;\; 
 \mprocU
 \;, \qquad
 \mathcal{M}^{ }_\rmi{b} 
 \; = \; 
 \hspace*{0.45cm}
 \mprocR
 \;, \qquad
 \mathcal{M}^{ }_\rmi{c} 
 \; = \; 
 \hspace*{0.45cm}
 \mprocS
 \;, 
 \\
 \mathcal{M}^{ }_\rmi{d} 
 & = & 
 \hspace*{0.45cm}
 \mprocN
 \;\; + \;\; 
 \mprocO
 \;, \qquad
 \mathcal{M}^{ }_\rmi{e} 
 \; = \; 
 \hspace*{0.45cm}
 \mprocP
 \;\; + \;\; 
 \mprocQ
 \;, 
 \\
 \mathcal{M}^{ }_\rmi{f} 
 & = & 
 \hspace*{0.45cm}
 \mprocA
 \;\; + \;\; 
 \mprocB
 \;\; + \;\; 
 \mprocC
 \;\; + \;\; 
 \mprocD
 \;, 
 \\ 
 \mathcal{M}^{ }_\rmi{g} 
 & = & 
 \hspace*{0.45cm}
 \mprocE
 \;, \qquad
 \mathcal{M}^{ }_\rmi{h} 
 \; = \; 
 \hspace*{0.45cm}
 \mprocG
 \;\; + \;\; 
 \mprocH
 \;\; + \;\; 
 \mprocI
 \;, 
 \\
 \mathcal{M}^{ }_\rmi{i} 
 & = & 
 \hspace*{0.45cm}
 \mprocJ
 \;\; + \;\; 
 \mprocK
 \;, \qquad
 \mathcal{M}^{ }_\rmi{j} 
 \; = \; 
 \hspace*{0.45cm}
 \mprocL
 \;\; + \;\; 
 \mprocM
 \;. 
\end{eqnarray*}

\caption[a]{\small 
 $2\to 2$ annihilation processes 
 leading to the coefficients in \eq\nr{c1}. Thick solid lines
 stand for Majorana particles, thick dashed lines for coloured scalars, 
 wiggly lines for gluons, arrowed lines for quarks, and 
 thin dashed lines for Higgs bosons and longitudinal polarizations of
 $W^\pm$ and $Z^0$ bosons. 
} 
\la{fig:processes}
\end{figure}
%

The first step is to determine annihilation cross
sections for all possible processes with dark matter initial states. 
The leading order Feynman diagrams are shown in 
\fig\ref{fig:processes}. According to the optical theorem, 
the amplitudes squared $|\mathcal{M}|^2$ can be expressed as
an imaginary (or ``absorptive'') contribution to an effective
Lagrangian~\cite{bodwin}. 

An important simplification in the Majorana case follows from 
the identity satisfied by Pauli matrices, 
$ 
 \sigma^k_{pq} \sigma^k_{rs}
 = 2 \delta^{ }_{ps}\delta^{ }_{qr} - \delta^{ }_{pq}\delta^{ }_{rs}
$.
Therefore a possible spin-dependent operator can be reduced to 
a spin-independent one:
$ \displaystyle
 \psi^\dagger_p \psi^\dagger_r \psi^{ }_s \psi^{ }_q 
 \, \sigma^k_{pq} \sigma^k_{rs}  
 = 
 - 3 
 \psi^\dagger_p \psi^\dagger_q \psi^{ }_q \psi^{ }_p 
$. 

At leading order in an expansion in $1/M^2$, 
the absorptive operators read
\ba
 \mathcal{L}^{ }_\rmi{abs} & = & 
 i \, \Bigl\{ 
 c^{ }_1 \, 
 \psi^\dagger_p \psi^\dagger_q \psi^{ }_q \psi^{ }_p 
 + 
 c^{ }_2 \, 
 \bigl(
   \psi^\dagger_p \phi^\dagger_\alpha \psi^{ }_p \phi^{ }_\alpha +  
   \psi^\dagger_p \varphi^\dagger_\alpha \psi^{ }_p \varphi^{ }_\alpha   
 \bigr)
 \nn 
 & + & 
 c^{ }_3 \,  
 \phi^\dagger_\alpha \varphi^\dagger_\alpha \varphi^{ }_\beta \phi^{ }_\beta
 + 
 c^{ }_4 \,  
 \phi^\dagger_\alpha  \varphi^\dagger_\beta\,
 \varphi^{ }_\gamma \phi^{ }_\delta
 \, T^{a}_{\alpha\beta} T^{a}_{\gamma\delta}
  + 
 c^{ }_5 \, 
 \bigl( \phi^\dagger_\alpha \phi^\dagger_\beta
        \phi^{ }_\beta \phi^{ }_\alpha 
   +    \varphi^\dagger_\alpha \varphi^\dagger_\beta
        \varphi^{ }_\beta \varphi^{ }_\alpha 
 \bigr)
 \Bigr\} 
 \;. \la{Labs}
\ea
Here $T^a$ are Hermitean generators of SU($\Nc$). 
In the partial wave language,
the operators in \eq\nr{Labs} correspond to $s$-wave 
annihilations, whereas $p$-wave annihilations would lead
to operators of $\rmO(1/M^4)$. 
At leading order in couplings, the coefficients read 
\ba
 && 
 c^{ }_1 \; = \; 0 
 \;, \quad 
 c^{ }_2 \; = \;
 \frac{|y|^2
 (
   |h|^2 + g_s^2 \CF^{ }
 ) 
 }{128\pi M^2}
 \;, \nn 
 && 
 c^{ }_3 \; = \;
 \frac{1}{32\pi M^2}
 \biggl(
  \lambda_3^2 + \frac{g_s^4 \CF^{ }}{\Nc} 
 \biggr)
 \;, \quad 
 c^{ }_4 \; = \;
 \frac{ g_s^4 (\Nc^2 - 4) }{64\pi M^2 \Nc^{ }}
 \;, \quad 
 c^{ }_5 \; = \;
 \frac{|y|^4}{128\pi M^2}
 \;. \la{c1}
\ea
A non-zero value of $c^{ }_1$ may be generated at higher orders. 
To minimize the magnitude of higher-order effects,
the couplings should be 
evaluated at the $\msbar$ renormalization scale $\bmu \sim 2M$.
We note that $c^{ }_5$ gets contributions from 
the ``Majorana-like'' processes
$\mathcal{M}^{ }_\rmi{i}$ and $\mathcal{M}^{ }_\rmi{j}$
in \fig\ref{fig:processes}, but not from
the ``Dirac-like'' amplitude $\mathcal{M}^{ }_\rmi{h}$.

%
\subsection{Number density, effective cross section, evolution equation}
\la{se:splitting}

Within Boltzmann equations
the overall dark matter abundance evolves as~\cite{clas1,clas2,old1} 
\be
 \dot{n} \; = \; -\langle \sigma^{ }_\rmi{eff}\, v \rangle \, 
 \bigl( n^2 - n_\rmi{eq}^2 \bigr) 
 \;, \la{boltzmann}
\ee
where $\dot{n}$ is the covariant time derivative in an expanding background.
To go beyond the quasiparticle approximation underlying the Boltzmann
approach, the effective cross section can be re-interpreted as 
a chemical equilibration rate, $\Gamma^{ }_\rmi{chem}$,  
and then defined on the 
non-perturbative level within linear response theory~\cite{chemical}. 
Furthermore, within the non-relativistic effective theory, 
$\Gamma^{ }_\rmi{chem}$
can be related to the thermal expectation value of 
$\mathcal{L}^{ }_\rmi{abs}$ from \eq\nr{Labs}~\cite{4quark_lattice}. 
These relations can be expressed as 
\be
 \langle \sigma^{ }_\rmi{eff}\, v \rangle
  \; = \; \frac{\Gamma^{ }_\rmi{chem}}{2 n^{ }_\rmi{eq}}
  \; = \; \frac{4}{n_\rmi{eq}^2}
  \langle \im \mathcal{L}^{ }_\rmi{abs} \rangle
 \;. \la{sigma}
\ee
In our model the number density amounts to 
\be
 n^{ }_\rmi{eq} \; = \; 
 \int_{\vec{p}} e^{-E^{ }_p/T}
 \, 
 \Bigl\{
  2 + 2 \Nc^{ }\, e^{-\Delta M^{ }_T / T} 
 \Bigr\}
 \;, \quad
 E^{ }_p \equiv M + \frac{\vec{p}^2}{2 M}
 \;. 
 \la{neq}
\ee
The mass difference $ \Delta M^{ }_T $ 
gets a vacuum contribution, 
$
 \Delta M =  M^{ }_\eta - M
$, 
and a thermal correction from \eq\nr{htl_re} as well as from 
a similar tadpole involving $\lambda^{ }_3$, 
\be
 \Delta M^{ }_T \; \equiv \; 
 \Delta M 
 + \frac{(g_s^2 \CF^{ } + \lambda^{ }_3) T^2}{12 M}
 - \frac{g_s^2 \CF^{ } \mD^{ }}{8\pi}
 \;. \la{D_M_T}
\ee
Note that the negative Salpeter correction may  
cancel against the positive terms.
At leading order the Debye mass parameter amounts to 
\be
 \mD^{ }= g_s^{ } T \, \sqrt{\frac{\Nc}{3} + \frac{\Nf}{6}}
 \;, 
\ee
where $\Nf$ is the number of quark flavours
(cf.\ ref.~\cite{Ghi-Sch} for higher orders). 
The effective values of $g^{ }_s$ and $\Nf$ 
are changed with the temperature, as reviewed in appendix~A.

For future reference we define a ``tree-level'' effective cross section, 
$ 
  \langle \sigma^{ }_\rmi{eff}\, v \rangle^{ (0)}
$, 
by evaluating the thermal expectation value 
$
 \langle \im \mathcal{L}^{ }_\rmi{abs} \rangle
$
at leading order and then making use of \eqs\nr{sigma} and \nr{neq}. 
Wick contracting the indices in \eq\nr{Labs} leads to 
\be
 \bigl\langle \sigma^{ }_\rmi{eff}\, v \bigr\rangle^{ (0)}
 \; = \; 
 \frac{
 2 c^{ }_1
 + 
 4 c^{ }_2 \Nc^{ }\, e^{-\Delta M^{ }_T / T} 
 + 
 [c^{ }_3 + c^{ }_4 \CF^{ } + 2 c^{ }_5 (\Nc^{ }+1)]
 \Nc^{ }\, e^{-2 \Delta M^{ }_T / T}
 }
 {\bigl(1 + \Nc^{ }\, e^{-\Delta M^{ }_T/T} \bigr)^2}
 \;. \la{sigma_0}
\ee

%
\subsection{Plasma-modified Schr\"odinger equation and 
generalized Sommerfeld factors}

Going beyond leading order, we evaluate 
$
 \langle \im \mathcal{L}^{ }_\rmi{abs} \rangle
$
as elaborated upon in ref.~\cite{threshold}, expressing it as
a Laplace transform of a spectral function characterizing the dynamics 
of the dark matter particles before their annihilation.
Denoting by~$E'$ the energy of the relative motion 
and by~$\vec{k}$ the momentum of the center-of-mass motion, this implies 
\ba
 \bigl\langle \im \mathcal{L}^{ }_\rmi{abs} \bigr\rangle
  & \approx &
 \int_{\vec{k}} e^{-\frac{2M}{T} - \frac{k^2}{4 M T}}
 \int_{-\Lambda}^{\infty} \! \frac{{\rm d} E'}{\pi} \, e^{-E'/T}
 \; \sum_i c^{ }_i\, \rho^{ }_i(E') 
 \nn
 & = & 
 \Bigl( \frac{MT}{\pi} \Bigr)^{3/2} e^{-2M/T}
 \int_{- \Lambda}^{\infty} 
 \! \frac{{\rm d}E'}{\pi} \, e^{-E'/T}
 \;  \sum_i c^{ }_i\, \rho^{ }_i(E')
 \;, \la{Laplace}
\ea 
where $\alpha^2 M \ll \Lambda  \ll M$ restricts the average
to the non-relativistic regime.\footnote{%
 Some elaboration about the need to introduce such a cutoff can be
 found in ref.~\cite{threshold}.
 In practice, we choose $\Lambda \simeq 2 \alpha^2 M$,
 and have verified that making it e.g.\ 2-3 times larger
 plays no role on our numerical resolution.
}
The spectral functions are obtained 
as imaginary parts of Green's functions,\footnote{%
 At higher orders in the non-relativistic expansion, kinetic terms and
 potentials suppressed by powers of $1/M^2$ could be added. In addition,
 operators suppressed by $1/M^4$ should be added in \eq\nr{Labs}.
} 
\ba
 \biggl[ 
   -\frac{\nabla_r^2}{M} + \mathcal{V}^{ }_i(r) - E'
 \biggr] G^{ }_i(E';\vec{r},\vec{r'}) & = & 
 N^{ }_i\, \delta^{(3)}(\vec{r}-\vec{r'})
 \;, \la{Seq} \\ 
 \lim_{\vec{r,r'}\to \vec{0}} \im  G^{ }_i(E';\vec{r},\vec{r'})
 & = & \rho^{ }_i(E') \la{get_rho}
 \;. \la{rho_def}
\ea
Here $\mathcal{V}^{ }_i$ contains a negative imaginary part, 
and $N^{ }_i$ is a normalization factor giving the number of 
contractions related to the operator that $c^{ }_i$ multiplies
in \eq\nr{Labs}:
\be
 N^{ }_1 \;\equiv\; 2 \;, \quad
 N^{ }_2 \;\equiv\; 4 \Nc^{ } \;, \quad
 N^{ }_3 \;\equiv\; \Nc^{ } \;, \quad
 N^{ }_4 \;\equiv\; \Nc^{ }\CF^{ } \;, \quad
 N^{ }_5 \;\equiv\; 2 \Nc^{ } (\Nc^{ } + 1 ) 
 \;. \la{Ns}
\ee

If the potentials $\mathcal{V}^{ }_i(r)$ were
$r$-independent and with an infinitesimal imaginary part, 
i.e.\ $\mathcal{V}^{ }_i(r) = \re \mathcal{V}^{ }_i(\infty) - i 0^+$, 
they would only induce mass shifts. 
In this case the spectral functions can be determined analytically, 
\be
 \rho^{(0)}_i(E') =  \frac{ N^{ }_i M^{\fr32} }{ 4\pi } 
 \, \theta\bigl( E' - \re \mathcal{V}^{ }_i(\infty) \bigr) 
 \, \sqrt{E' - \re \mathcal{V}^{ }_i(\infty) } 
 \;. 
\ee
This form can be used for defining generalized Sommerfeld factors: 
\be
 \bar{S}^{ }_i \; \equiv \; 
 \frac{
 \int_{- \Lambda}^{\infty} 
 \! \frac{{\rm d}E'}{\pi} \, e^{-E'/T}
 \, \rho^{ }_i(E')
 }
 {
 \int_{- \Lambda}^{\infty} 
 \! \frac{{\rm d}E'}{\pi} \, e^{-E'/T}
 \, \rho^{(0)}_i(E')
 }
 \; = \; 
 \Bigl(\frac{4\pi}{MT} \Bigr)^{\fr32}
 \int_{- \Lambda}^{\infty} 
 \! \frac{{\rm d}E'}{\pi} 
 \, e^{[ {\rm Re} \mathcal{V}^{ }_i(\infty)-E'] / {T}}
 \, \frac{ \rho^{ }_i(E') }{N^{ }_i} 
 \;. \la{barS}
\ee
Then \eq\nr{Laplace} combined with 
\eqs\nr{sigma} and \nr{neq} leads to a generalization of \eq\nr{sigma_0}, 
\be
 \bigl\langle \sigma^{ }_\rmi{eff}\, v \bigr\rangle^{ }
 \; = \; 
 \frac{
 2 c^{ }_1
 + 
 4 c^{ }_2 \Nc^{ }\, e^{-\Delta M^{ }_T / T} 
 + 
 [c^{ }_3 \bar{S}^{ }_3
 + c^{ }_4 \bar{S}^{ }_4 \CF^{ } 
 + 2 c^{ }_5 \bar{S}^{ }_5 (\Nc^{ }+1)]
 \Nc^{ }\, e^{-2 \Delta M^{ }_T / T}
 }
 {\bigl(1 + \Nc^{ }\, e^{-\Delta M^{ }_T/T} \bigr)^2}
 \;. \la{sigma_final}
\ee

If a potential $\mathcal{V}^{ }_i(r)$ leads to a bound state, 
whose width is much smaller than the binding energy, the corresponding
generalized Sommerfeld factor can be computed in analytic form. In this 
case \eq\nr{Seq} can be solved in a spectral representation, 
resulting in 
\be
 \rho^{ }_i (E') = \pi N^{ }_i \sum_j 
 \frac{|\psi^{ }_j(\vec{0})|^2\,
 \delta(E' - E'_{j})
 }
 {\int \! {\rm d}^3\vec{r}\, |\psi^{ }_j(\vec{r}) |^2 } 
 \;, 
\ee
where $\psi^{ }_j$ are the bound state wave functions. 
Inserting into \eq\nr{barS}, 
the contribution
of the $j$th bound state to $\bar{S}^{ }_i$ reads 
\be
 \Delta^{ }_j \bar{S}^{ }_i 
 = 
 \Bigl(\frac{4\pi}{MT} \Bigr)^{\fr32}
 \, \frac{|\psi^{ }_j(\vec{0})|^2
 \, e^{[ {\rm Re} \mathcal{V}^{ }_i(\infty)-E'_j] / {T}} }
 {\int \! {\rm d}^3\vec{r}\, |\psi^{ }_j(\vec{r}) |^2 } 
 \;. \la{barSij}
\ee
This becomes (exponentially) large 
when $T \ll \alphas^2 M$, however chemical equilibrium is 
lost in the dark sector at low $T$, which imposes an 
effective cutoff on the growth 
(cf.\ \ses\ref{se:numerics} and \ref{se:concl}). 

%
\subsection{Thermal potentials}

In order to write down the potentials $\mathcal{V}^{ }_i(r)$ appearing
in \eq\nr{Seq}, let us define
\ba
 {v}^{ }_\rmii{ }(r) & \equiv &   
 \frac{g_s^2}{2} 
 \int_{\vec{k}} e^{i \vec{k}\cdot\vec{r}}
 \, \biggl\{ 
 \frac{1}{k^2 + \mD^2}
 \; - \; \frac{i\pi T}{k}
 \frac{\mD^2}
        {(k^2 + \mD^2)^2}
 \biggr\}
 \la{vr_def} \\[3mm] 
 & = & 
 \frac{g_s^2}{2} \times 
 \left\{
 \begin{array}{ll}
 \displaystyle
 \frac{\exp({-\mD^{ }r})}{4\pi r}
 - 
 \frac{i T}{2 \pi \mD^{ } r}
 \int_0^\infty \! \frac{{\rm d} z \, \sin(z \mD^{ }r)}{(z^2 +1)^2}
   & \;, \quad r > 0 \\[3mm]
 \displaystyle 
 -\frac{\mD^{ }}{4\pi} 
 - \frac{i T}{4\pi}
   & \;, \quad r = 0 
 \end{array}
 \right.
 \;. 
 \la{vr}
\ea 
The integrand in \eq\nr{vr_def}
corresponds to the static limit of the time-ordered HTL-resummed
temporal gluon propagator. Then we find
\ba
 \mathcal{V}^{ }_1(r) 
 & = &  
 0 
 \;, \quad
 \mathcal{V}^{ }_2(r) 
 \; = \; 
 \CF^{ }\, v(0) 
 \;, \quad
 \mathcal{V}^{ }_{3}(r)
 \; = \; 
   2 \CF^{ }\, \bigl[ v(0) - v(r) \bigr]
 \;, \la{V3} \\[3mm]
 \mathcal{V}^{ }_{4}(r)
 & = & 
   2 \CF^{ }\, v(0) + \frac{v(r)}{\Nc^{ }}  
 \;, \quad
 \mathcal{V}^{ }_{5}(r)
 \; = \;  
   2 \CF^{ }\, v(0) + \frac{(\Nc^{ } - 1)v(r)}{\Nc^{ }} 
 \;. \la{V5} 
\ea
The structure $\mathcal{V}^{ }_3(r)$ equals the combination 
$V(r) - i \Gamma(r)$ shown in \eqs\nr{ex_G}--\nr{ex_Gamma}, 
whereas $\CF^{ } \re[ v(0) ]$ yields the Salpeter part of $\Delta M^{ }_T$
in \eq\nr{D_M_T}. 
The potential $\mathcal{V}^{ }_3(r)$ corresponds to a singlet
potential, $\mathcal{V}^{ }_4(r)$ to an octet potential, and 
$\mathcal{V}^{ }_5(r)$ to a particle-particle potential, 
relevant because of the presence of a 
particle-particle annihilation channel generated by Majorana exchange
(cf.\ the discussion around the end of \se\ref{ss:im}).

We note in passing that at $T < 160$~GeV, when the Higgs mechanism
is operative, additional potentials can be generated, particularly
through the Higgs portal coupling $\lambda^{ }_3$ in \eq\nr{L}
(cf.\ e.g.\ ref.~\cite{hp}). However
the coefficients of these potentials are suppressed by 
$\sim \lambda_3^2 v^2/M^2$, where
$v$ is the Higgs expectation value. 
Given that we consider $M \ge 2$~TeV, 
we expect their contributions to 
be negligible compared with QCD effects and have not included them. 
We also note that an $r$-dependence  
can be generated for $\mathcal{V}^{ }_2(r)$ 
through quark exchange, however this is suppressed by 
$\sim |y|^2 \mathbf{\sigma}\cdot\mathbf{\nabla} / M$.

For a practical use of \eq\nr{vr}, numerical values are needed
for the parameters $g_s^2$ and $\mD^2$. We relegate a discussion
of this point into appendix~A. Let us however note 
that we restrict to temperatures
$T \gsim 1$~GeV, so that the real part of the potential contains no 
trace of a string tension~\cite{bkr}. 
Furthermore, in accordance with 
the low-temperature gluon-radiation
contribution specified below \eq\nr{ex_Gamma} and with 
general arguments presented in ref.~\cite{idm}, the imaginary
part of the potential is multiplied by the Boltzmann factor
$e^{-|E'|/T}$ for $E' < 0$.

%
\section{Numerical evaluations}
\la{se:numerics}

\begin{figure}[t]

\hspace*{-0.1cm}
\centerline{%
 \epsfxsize=7.6cm\epsfbox{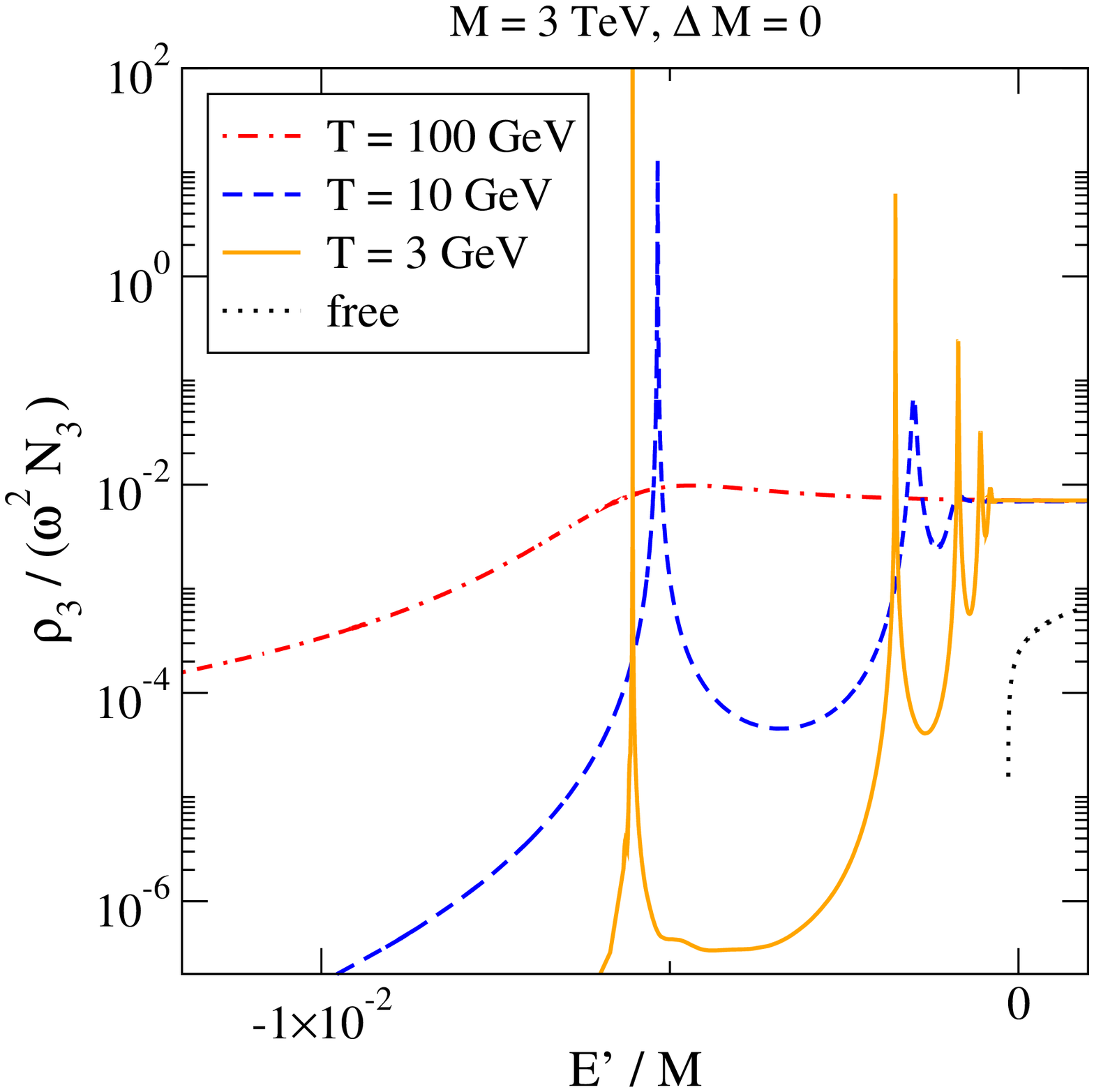}
 \hspace{0.1cm}
 \epsfxsize=7.6cm\epsfbox{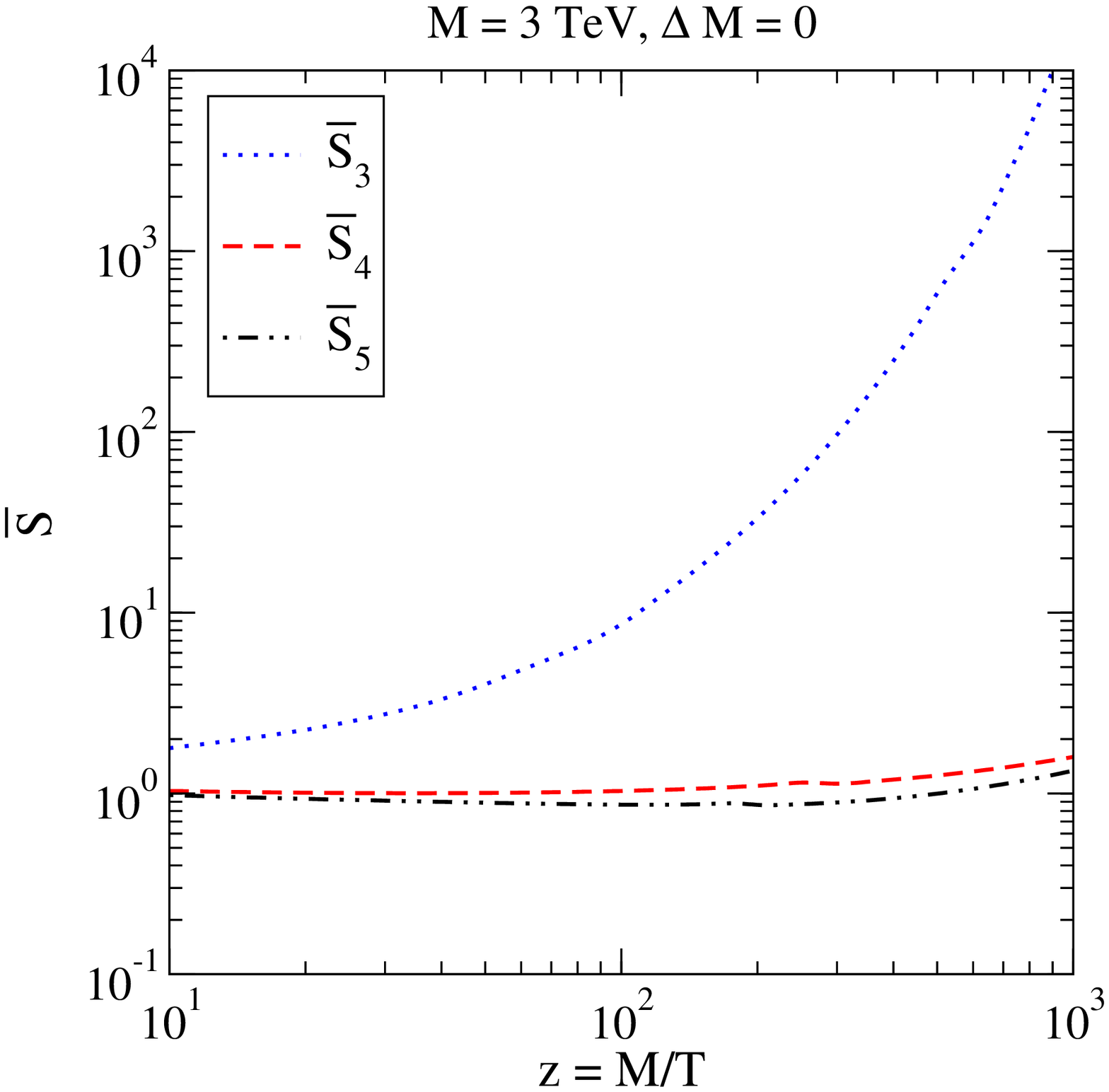}%
}

\caption[a]{\small
 Left: 
 the spectral function $\rho^{ }_3$ of the scalar pair, 
 interacting via the attractive potential $\mathcal{V}^{ }_3$.
 Here $\omega \equiv 2 M + E'$.
 At low temperatures a dense spectrum of bound states
 can be observed, which gradually ``melts away'' as the temperature
 increases. 
 Right: 
 The generalized Sommerfeld factors, \eq\nr{barS}, 
 corresponding to the annihilation of the coloured 
 scalars via different channels. 
}

\la{fig:M3000}
\end{figure}

\begin{figure}[t]

\hspace*{-0.1cm}
\centerline{%
 \epsfxsize=7.6cm\epsfbox{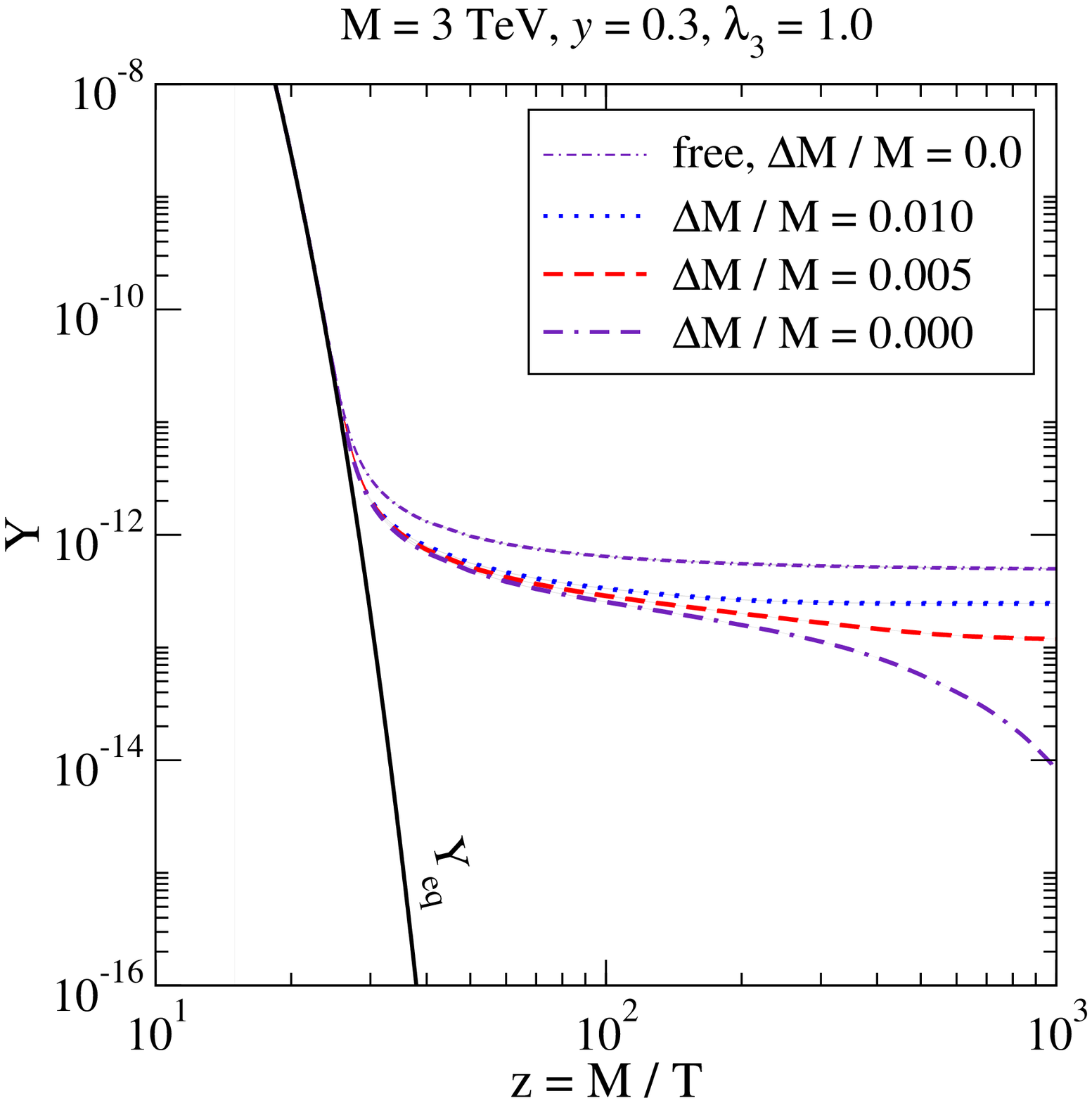}
 \hspace{0.1cm}
 \epsfxsize=7.6cm\epsfbox{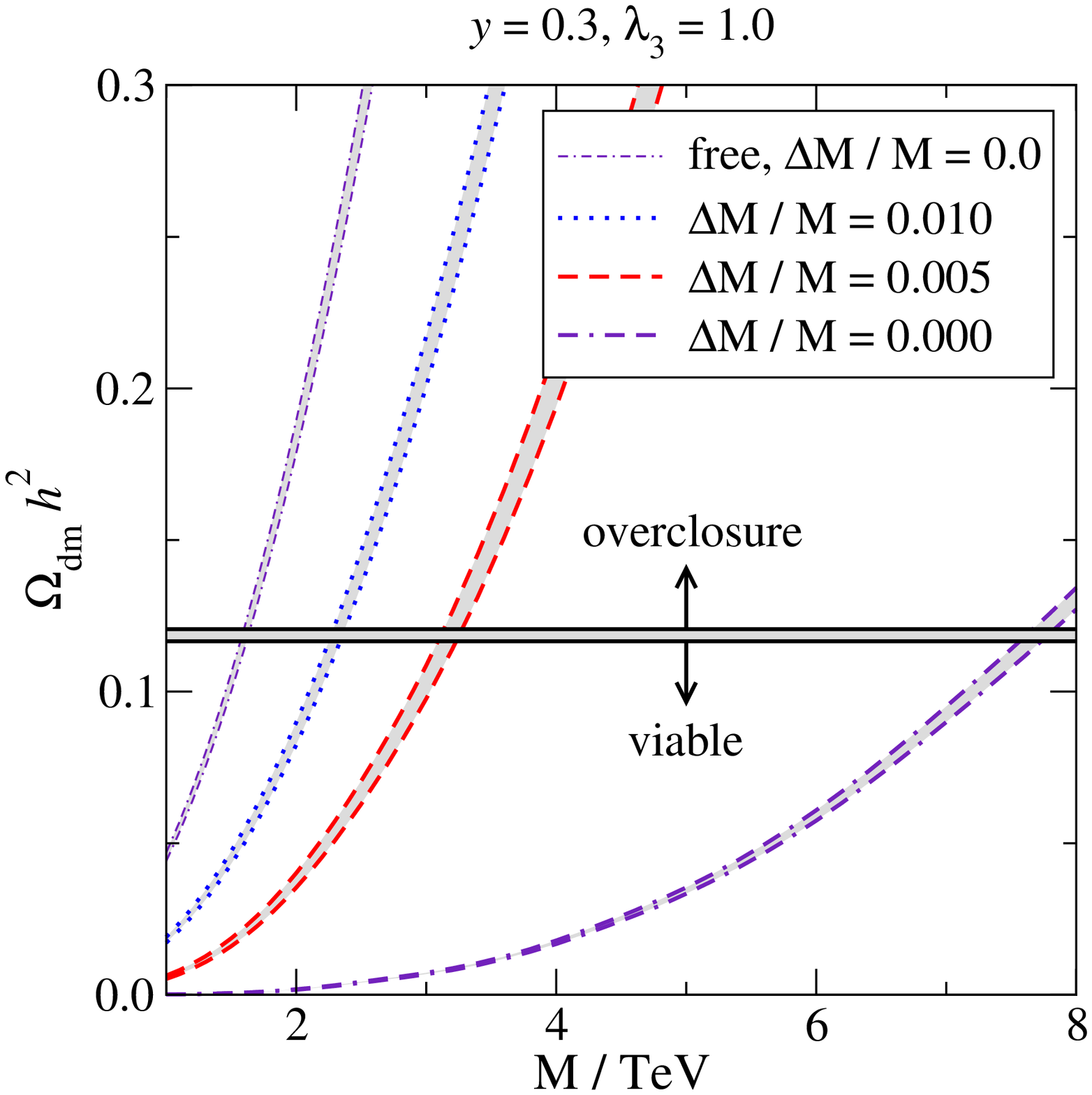}%
}

\caption[a]{\small
 Left: 
 solutions of \eq\nr{dY} for $M = 3$~TeV and selected values
 of $\Delta M/M$. 
 The quark Yukawa coupling $h$ is either $0.0$ (upper edges of bands) or
 $1.0$ (lower edges). 
 If $\Delta M/M$ is too small, 
 dark matter may convert to coloured scalars and 
 get efficiently annihilated; this is only partly visible, 
 because we have stopped the integration 
 at $z^{ }_\rmi{final} = 10^3$.  
 Right: 
 the dark matter abundance at $z^{ }_\rmi{final} = 10^3$.
 The horizontal band shows the observed value
 $ 
  0.1186(20)
 $~\cite{planck}. 
}

\la{fig:Y}
\end{figure}

\begin{figure}[t]

\hspace*{-0.1cm}
\centerline{%
 \epsfxsize=7.6cm\epsfbox{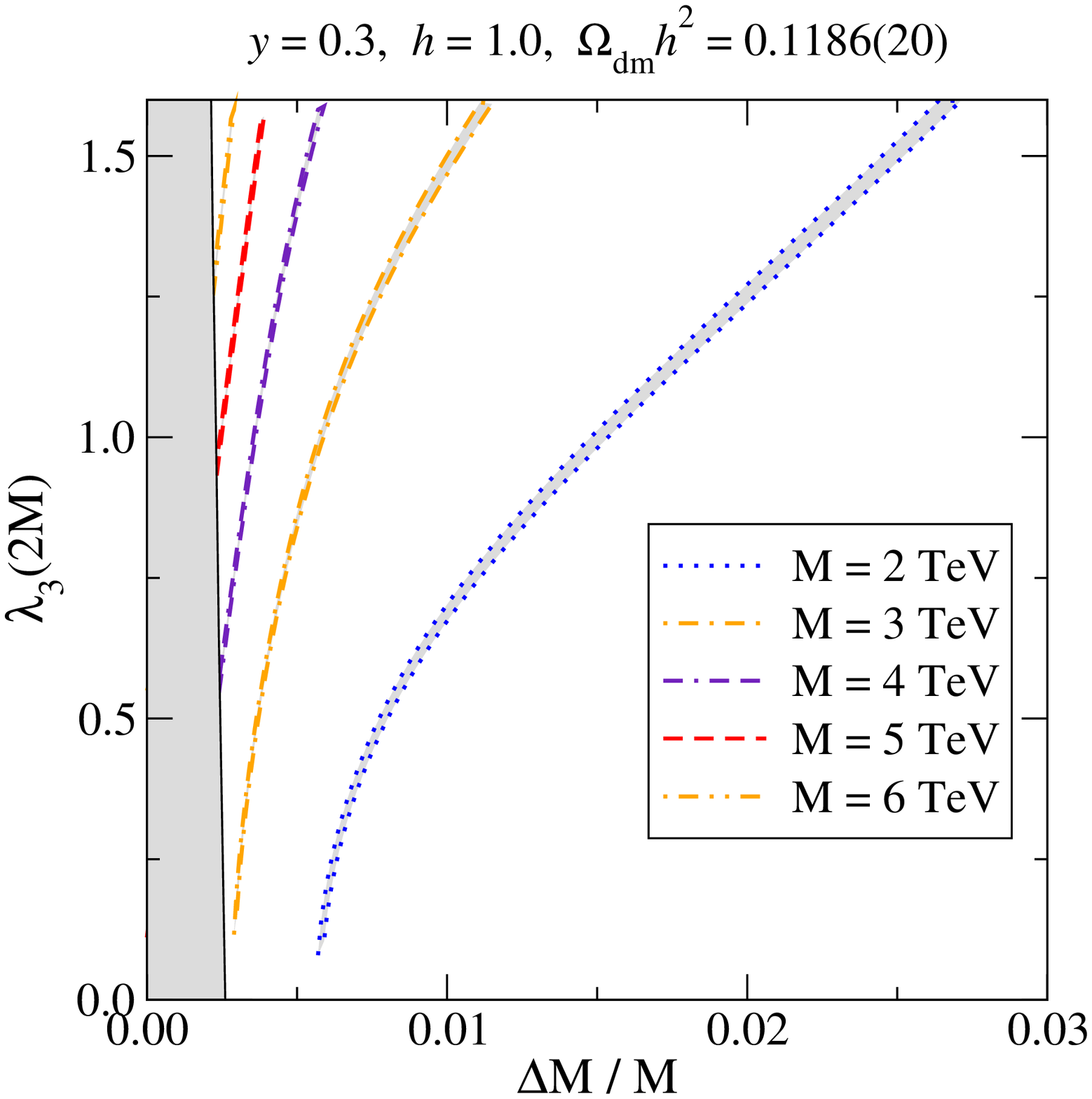}
 \hspace{0.1cm}
 \epsfxsize=7.6cm\epsfbox{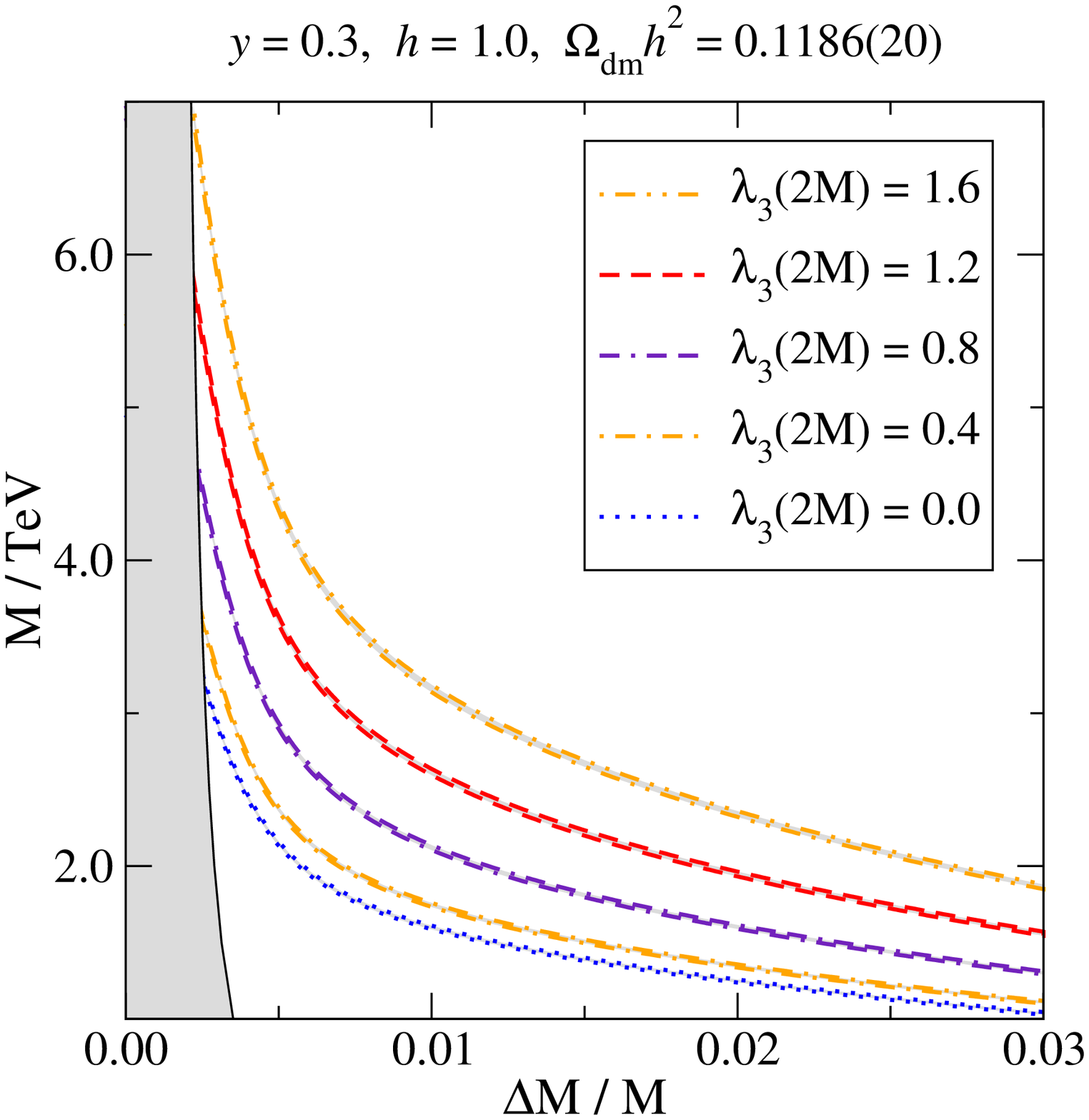}
}

\caption[a]{\small
 Left:
 values of the coupling $\lambda^{ }_3(2M)$ 
 needed for producing the correct dark matter abundance. 
 In the shaded region, 
 bound states of coloured scalars are lighter 
 than scattering states of two Majorana fermions.  
 Right: 
 a corresponding plot for the dark matter 
 mass scale $M$/TeV. Note that $h$ has two different meanings here, 
 a quark Yukawa coupling and a rescaled Hubble rate. 
}

\la{fig:Omega}
\end{figure}

Having determined the spectral functions from \eqs\nr{Seq} and \nr{rho_def}
and the generalized Sommerfeld factors from \eq\nr{barS} or \nr{barSij},
the effective cross section is obtained from \eq\nr{sigma_final}. 
Subsequently \eq\nr{boltzmann} can be integrated for the dark 
matter abundance. As usual we define a yield parameter 
as $Y \equiv n/s$, where $s$ is the entropy density, and change
variables from time to $z \equiv M/T$, whereby \eq\nr{boltzmann} becomes
\be
 Y'(z) 
   = - \, 
   \langle \sigma^{ }_\rmi{eff}\, v \rangle M m^{ }_\rmi{Pl} \times 
   \frac{c(T)}{\sqrt{24\pi e(T)}} \times
   \left. \frac{Y^2(z) - Y^{2}_\rmi{eq}(z) }{z^2} \right|^{ }_{T = M/z}
 \;. \la{dY}
\ee
Here $m^{ }_\rmi{Pl}$ is the Planck mass, 
$e$ is the energy density, and $c$ is the heat capacity, 
for which we use values from ref.~\cite{crossover} 
(cf.\ also ref.~\cite{dono}).
The final value $ Y(z^{ }_\rmi{final}) $ 
yields the energy fraction  
$
   \Omega^{ }_\rmi{dm} h^2 =
           {Y(z^{ }_\rmi{final})\,M}/
           {[3.645 \times 10^{-9}\,\mbox{GeV}]}
$.

We integrate \eq\nr{dY} up to $z^{ }_\rmi{final} = 10^3$.
At around these temperatures, depending on the value of $\Delta M/M$, 
the processes of interest have either ceased to be active, or are 
falling out of chemical equilibrium, because their rates are suppressed by 
$e^{-\Delta M / T} \ll 1$. Therefore they cannot be 
reliably addressed within the current framework. 

In~\fig\ref{fig:M3000}(left) we show the spectral function $\rho^{ }_3$
corresponding to the attractive channel, displaying a dense spectrum of
bound states at low temperatures. The corresponding generalized 
Sommerfeld factor, obtained from \eq\nr{barS}, 
is shown in \fig\ref{fig:M3000}(right). 
An exponential increase is observed at low temperatures, as indicated
by \eq\nr{barSij}. The repulsive channels also show a modest increase
at very low temperatures, due to the fact that the spectral function
extends below the threshold at finite temperature~\cite{idm}. 
Examples of results
obtained by integrating \eq\nr{dY} are shown in \fig\ref{fig:Y}. 
In particular, it can be observed how a very efficient annihilation
sets in at low temperatures, if $\Delta M$ is small so that 
bound states of coloured scalars are lighter than scattering states
of Majorana fermions. Finally, \fig\ref{fig:Omega} shows slices
of the parameter space leading to the correct dark matter abundance. 

In the plots the Yukawa couplings have been set to 
the stop-like values $y=0.3, h=1.0$. However these couplings only
have a modest effect if chosen otherwise, because they do 
not affect the coefficient~$c^{ }_3$ appearing the attractive
channel, cf.\ \eq\nr{c1}. As an example, setting $h=0.0$
increases the abundance typically by $\sim 5$\%, 
cf.\ \fig\ref{fig:Y}.
The most important role
is played by the coupling $\lambda^{ }_3$. 
For $c^{ }_3$ this coupling
has been evaluated at the scale $\bmu = 2M$, whereas for collider phenomenology
its value at a scale $\bmu \sim \mH^{ }$ would be more relevant. 
The latter can be obtained from \eq\nr{lam3}, 
and is some tens of percent smaller than $\lambda^{ }_3(2M)$. 
We stress that, as shown by \eq\nr{lam3}, Yukawa couplings
always generate a non-zero value for $\lambda^{ }_3$ 
through renormalization group running.  

%
\section{Conclusions}
\la{se:concl}

We have investigated a simple extension of the Standard Model,
cf.\ \se\ref{se:model}, which has become popular as 
a prototypical fix to the increasingly stringent empirical
constraints placed 
on ``WIMP''-like frameworks. In this model
dark matter consists of a Majorana fermion, which only has a $p$-wave 
annihilation cross section at tree level, 
helping to respect experimental non-observation 
constraints from indirect detection. 
The Majorana fermion 
has a Yukawa interaction with a QCD-charged scalar field
(such as a right-handed stop or sbottom in the MSSM) 
and a Standard Model quark. 
For large masses and small mass splittings between 
the Majorana fermion
and 
the scalar field, 
the best sensitivity for discovering the 
Majorana fermion appears to be  
direct detection by XENON1T~\cite{giv}, enhanced by resonant
scattering off quarks through scalar exchange, 
even if interactions with top or bottom quarks are 
much less constrained than those with up or down quarks. 

Despite its simplicity, the model displays rich physics in the early  
universe. We have extended previous investigations~\cite{eoz,hhk,ipsv,giv,%
 ll,mrss,klz} 
by incorporating the full spectrum of thermally broadened bound states 
as well as the effect of soft 
$2\leftrightarrow 2$ scatterings. In general such scatterings 
dominate interaction rates at 
small mass splittings, because they are not phase-space suppressed
in the same way as 
$1\leftrightarrow 2$ scatterings are, cf. \se\ref{se:thermal}.  

The reason that the model leads to a viable 
cosmology is that at high temperatures dark matter 
annihilates efficiently through the scalar channel, 
guaranteeing that its overall abundance remains low.  
The fast 
annihilations proceed particularly 
through bound states formed by 
the scalars, cf.\ \fig\ref{fig:M3000}. 
As shown in \fig\ref{fig:Omega}, the model
can be phenomenologically viable for masses up to $M \sim 5...6$~TeV, 
provided that the mass splitting is small, $\Delta M / M < 5\times 10^{-3}$, 
and that the ``Higgs portal'' coupling $\lambda^{ }_3$ between the coloured
scalar and the Higgs doublet is substantial. We recall that in 
supersymmetric theories, $\lambda^{ }_3$ is proportional to the 
quark Yukawa coupling squared, $\lambda^{ }_3\sim |h|^2$, and therefore indeed
large if we identify the coloured scalar as a right-handed stop. Actually, 
similar arguments 
but a somewhat more complicated analysis are expected to apply
to a left-handed stop as well
(cf.\ e.g.\ ref.~\cite{sf}). 

We believe that the mass splitting should not be too small, however. 
The non-relativistic binding energy of the lightest bound state, $E'_1$, 
is negative. 
If it overcompensates for the mass difference, so that 
$2\Delta M + E'_1 < 0$, the lightest two-particle states in the dark sector
are the bound states formed by the coloured scalars. 
However these 
states are short-lived. Therefore it seems possible that (almost) 
all dark matter converts into the scalars and
gets subsequently annihilated, so that the model may not be viable
as an explanation for the observed dark matter abundance. 
This domain has been excluded through the grey bands
in \fig\ref{fig:Omega}. 
If we close eyes to this concern and assume that chemical equilibrium
is maintained, 
then the value of~$M$ could be substantially larger 
than in \fig\ref{fig:Omega}, 
for instance $M \sim 8$~TeV as shown in \fig\ref{fig:Y}, 
and even more if we integrate down to lower temperatures.

We end by remarking that the model 
contains two portal couplings, $\lambda^{ }_3$ and $y$. The roles that
these play are rather different. The value of 
$\lambda^{ }_3$ at the scale $\bmu = 2M$ influences 
the coefficient
$c^{ }_3$ which mediates the most efficient annihilations, 
cf.\ \eq\nr{c1}. In contrast $y$ affects the rate
of transitions between the Majorana fermions and coloured
scalars, cf.\ \eq\nr{htlDM}, as well as the running of $\lambda^{ }_3$, 
cf.\ \eq\nr{lam3}. As long as $y$ is not miniscule, so that 
the rate in \eq\nr{htlDM} remains in equilibrium, it has in practice
little influence on our main results in \fig\ref{fig:Omega}.

%
\section*{Acknowledgements}

This work was supported by the Swiss National Science Foundation
(SNF) under grant 200020-168988. 

%
\appendix
\renewcommand{\thesection}{Appendix~\Alph{section}}
\renewcommand{\thesubsection}{\Alph{section}.\arabic{subsection}}
\renewcommand{\theequation}{\Alph{section}.\arabic{equation}}

%
\section{Fixing of vacuum and thermal couplings}

We start by listing the 1-loop renormalization group (RG) 
equations satisfied by the model of \se\ref{se:model}.
Apart from the couplings shown in \eq\nr{L}, 
the Higgs self-coupling $\lambda^{ }_1$,
the Higgs mass parameter $\mu_\rmii{$H$}^2$, 
the weak and strong gauge couplings $g_w^2,g_s^2$, 
and the Yukawa coupling~$h$ associated with 
the quark flavour~$q$ appear. The 
hypercharge coupling is omitted for simplicity. The number
of colours is denoted by $\Nc^{ }=3$, and 
$\CF^{ }\equiv (\Nc^2 -1)/(2\Nc^{ })$, whereas 
$\nG^{ }= 3$, $\nS^{ } = 1$ and $\nW^{ } = 1$ refer to 
the numbers of fermion generations, strongly 
interacting scalar triplets, and weakly interacting
scalar doublets, respectively. 

Parametrizing the $\msbar$ renormalization scale $\bmu$ through 
\be
 t \; \equiv \; \ln \bmu^2
 \;, 
\ee
we find
\ba
 \partial^{ }_t \mu_\rmii{$H$}^2 & = & 
 \frac{1}{(4\pi)^2}
 \biggl\{
  \biggl[ 6 \lambda^{ }_1 - \frac{9 g_w^2}{4} + |h|^2 \Nc^{ } \biggr] 
  \mu_\rmii{$H$}^2
 + \lambda^{ }_3 \Nc^{ } M_\eta^2 
 \biggr\} 
 \;, \\ 
 \partial^{ }_t M_\eta^2 & = & 
 \frac{1}{(4\pi)^2}
 \biggl\{
    \Bigl[ 2 \lambda^{ }_2 (\Nc^{ }+1) - 3 g_s^2 \CF^{ }+ |y|^2 \Bigr]
    M_\eta^2 
   + 2 \lambda^{ }_3 \mu_\rmii{$H$}^2 - 2 |y|^2 M^2
 \biggr\} 
 \;, \\ 
 \partial^{ }_t M^2 & = & 
 \frac{1}{(4\pi)^2}
 \biggl\{
   |y|^2 \Nc^{ } M^2
 \biggr\} 
 \;, \\ 
 \partial^{ }_t \lambda^{ }_1 & = & 
 \frac{1}{(4\pi)^2}
 \biggl\{
   \biggl[ 12 \lambda^{ }_1 - \frac{9 g_w^2}{2} + 2 |h|^2 \Nc^{ } \biggr]
   \lambda^{ }_1
   + \, \frac{\lambda_3^2 \Nc^{ }}{2} 
   + \frac{9 g_w^4}{16} 
   - |h|^4 \Nc^{ }
 \biggr\} 
 \;, \\ 
 \partial^{ }_t \lambda^{ }_2 & = & 
 \frac{1}{(4\pi)^2}
 \biggl\{
   \Bigl[ 2 \lambda^{ }_2 (\Nc^{ }+4)
  - 6 g_s^2 \CF^{ } + 2 |y|^2 \Bigr]
   \lambda^{ }_2
 \nn 
   &  & \hspace*{1cm} + \, \lambda_3^2  
   + \frac{3 (\Nc^3 + \Nc^2 - 4 \Nc^{ }+2 ) g_s^4}{8 \Nc^2} 
   - |y|^4 
 \biggr\} 
 \;, \\ 
 \partial^{ }_t \lambda^{ }_3 & = & 
 \frac{1}{(4\pi)^2}
 \biggl\{
   \biggl[ 6 \lambda^{ }_1
   + 2 \lambda^{ }_2 (\Nc^{ }+1) 
   + 2 \lambda^{ }_3 
   - \frac{9 g_w^2}{4}
   - 3 g_s^2 \CF^{ }
   + |y|^2 
   + |h|^2 \Nc^{ } \biggr]
   \lambda^{ }_3
 \nn 
   &  & \hspace*{1cm}
   - \, 2 |h|^2 |y|^2 
    \biggr\} 
 \;, \la{lam3} \\ 
 \partial^{ }_t |y|^2 & = & 
 \frac{ |y|^2 }{(4\pi)^2}
  \biggl\{
   \frac{|y|^2(\Nc^{ }+3)}{2} + |h|^2 - 3 g_s^2 \CF^{ } 
  \biggr\} 
 \;, \\ 
 \partial^{ }_t |h|^2 & = & 
 \frac{ |h|^2 }{(4\pi)^2}
  \biggl\{
   \frac{|h|^2(2\Nc^{ }+3)}{2} + \frac{|y|^2}{2}
  - \frac{9 g_w^2}{4} - 6 g_s^2 \CF^{ } 
  \biggr\} 
 \;, \\ 
 \partial^{ }_t g_w^2 & = & 
 \frac{ g_w^4 }{(4\pi)^2}
  \biggl\{
   \frac{\nW^{ }}{6} + \frac{4 \nG^{ }}{3} - \frac{22}{3}
  \biggr\} 
 \;, \\ 
 \partial^{ }_t g_s^2 & = & 
 \frac{ g_s^4 }{(4\pi)^2}
  \biggl\{
   \frac{\nS^{ }}{6} + \frac{4 \nG^{ }}{3} - \frac{11\Nc^{ }}{3}
  \biggr\} \la{gS_run}
 \;.
\ea
We note in particular that a non-zero value 
is generated for $\lambda^{ }_3$
by the running induced by Yukawa couplings, cf.\ \eq\nr{lam3}.

The only coupling that we need at a scale $\bmu \ll M$ is the strong
coupling. Since it has a large influence, we evaluate it at 2-loop
level for $\bmu \le M$ (nowadays running is known up to 5-loop 
level~\cite{alph1,alph2,alph3}). Denoting by $\Nf^{ }$ the 
number of flavours and
setting $\Nc^{ }= 3$ for brevity, the 2-loop running is given by
\ba
 \partial^{ }_t a^{ }_s & = & - \bigl\{ 
 \beta^{ }_0 a_s^2 + \beta^{ }_1 a_s^3 + \ldots 
 \bigr\}
 \;, \\[3mm] 
  a^{ }_s & \equiv & \frac{g_s^2}{4\pi^2}
 \;, \quad
 \beta^{ }_0 = \frac{1}{4} 
 \biggl\{
   11 - \frac{2 \Nf^{ }}{3} 
 \biggr\}
 \;, \quad
 \beta^{ }_1 = \frac{1}{4^2} 
 \biggl\{
   102 - \frac{38 \Nf^{ }}{3} 
 \biggr\}
 \;. 
\ea
The value of $\Nf^{ }= 3,...,6$  is changed when a quark mass threshold is
crossed at $\bmu = m^{ }_i$, where continuity is imposed. 
The initial value 
is $\alphas(\mZ^{ }) \simeq 0.118$. For $\bmu > M$, the contribution
of the coloured scalar is added and we switch over to 1-loop 
running, i.e.\ \eq\nr{gS_run}. 

When we evaluate the static potential,  
a wide range of distance scales appears. 
At short distances, inspired by refs.~\cite{pot1,pot4},
we evaluate the 2-loop coupling at the scale 
$\bmu = e^{-\gammaE} / r $. Since parametrically only the 
scales $\alpha M \ll M$ play a role in the Schr\"odinger equation, 
the running does not
include the coloured scalar in this domain.

At large distances, we employ effective thermal couplings. 
In the absence of NLO computations 
for thermal quarkonium observables, we adopt 
effective couplings from another context, that 
of dimensionally reduced field theories~\cite{dr1,dr2}. 
There the Debye mass parameter and an ``electrostatic''
coupling are expressed as~\cite{pheneos} 
\be
 \mD^2 \; \equiv \;
  T^2 \Bigl[ \, g_s^2\,  \aEms{4}
  + \frac{g_s^4}{(4\pi)^2}\, \aEms{6}
  + \rmO(g_s^6)\, \Bigr]
 \;, \quad
 g_\rmii{E}^2 \; \equiv \; 
 g_s^2 
  +  \frac{g_s^4}{(4\pi)^2}\, \aEms{7}
  + \rmO(g_s^6)
 \;. \la{gE}
\ee
For general masses, only $ \aEms{4} $ and $ \aEms{7} $ are 
available at present: 
\ba
 \aEms{4} \!\! & = & \!\!
 \frac{\Nc^{ }}{3} \; + \; 
  4 \sum_{i=1}^{\Nf}
 \biggl[
  F_2^{ }\biggl( \frac{m_i^2}{T^2}, 0 \biggr)
  - \frac{m_i^2}{(4\pi)^2T^2} \, 
  F_3^{ }\biggl( \frac{m_i^2}{T^2}, 0 \biggr)
  \biggr]
 \;, \la{aE4} \\
 \aEms{7} \!\! & = & \!\!
 \frac{22\Nc^{ }}{3}
 \biggl[
   \ln\biggl( \frac{\bmu e^{\gammaE}}{4\pi T} 
   \biggr) + \frac{1}{22}
 \biggr] 
 -  
 \fr23 \sum_{i=1}^{\Nf}
 \biggl[
  \theta(\bmu - m^{ }_i)\, \ln \biggl( \frac{\bmu^2}{m_i^2} \biggr) + 
  F_3\biggl( \frac{m_i^2}{T^2},0 \biggr)
  \biggr]
 \;. \hspace*{1cm} \la{aE7}
\ea
Here the functions read
($\hat{n}_\rmii{F}^{ }(x) \equiv 1/(e^x + 1)$; chemical potentials
have been set to zero)
\ba
  F_2(y,0) \!\! & \equiv & \!\! 
 \frac{1}{4\pi^2} \int_0^\infty \! {\rm d} x\, 
 \biggl[ \frac{x}{x+y} \biggr]^{\fr12}
   \, \hat{n}_\rmii{F}^{ }\bigl(\sqrt{x+y}\bigr) 
 \; = \; 
 \frac{1}{24} + \rmO(y)
 \;, \la{F2_def} \\ 
 F_3(y,0) \!\! & \equiv & \!\! 
 - 2 \int_0^\infty \! {\rm d} x\, 
 \biggl[ \frac{x}{x+y} \biggr]^{\fr12}
 \frac{ \hat{n}_\rmii{F}^{ }\bigl(\sqrt{x+y}\bigr) }{x}
 \; = \; 
 \ln \Bigl( \frac{y}{\pi^2} \Bigr) + 2 \gammaE + \rmO(y)
 \;. \la{F3_def}
\ea
Given that $\aEms{6}$ is not currently known for general masses, 
we estimate 
\be
 \mD^2 \simeq T^2 g_\rmii{E}^2\, \aEms{4} 
 \;, 
\ee
inserting here PDG values for the quark masses~\cite{pdg}. The scale
parameter is set to $\bmu = 2\pi T$.

\small{
%

}


\begin{thebibliography}{99}

\bibitem{eg}
  J.~Edsj\"o and P.~Gondolo,
  {\it Neutralino relic density including coannihilations,}
  Phys.\ Rev.\ D {56} (1997) 1879
  [hep-ph/9704361].

\bibitem{old32}
  W.~Detmold, M.~McCullough and A.~Pochinsky,
  {\it Dark Nuclei I: Cosmology and Indirect Detection,}
  Phys.\ Rev.\ D {90} (2014) 115013
  [1406.2276].

\bibitem{old4}
  B.~von Harling and K.~Petraki,
  {\it Bound-state formation for thermal relic dark matter and unitarity,}
  JCAP {12} (2014) 033
  [1407.7874].

\bibitem{4quark_lattice}
  S.~Kim and M.~Laine,
  {\it Rapid thermal co-annihilation through bound states in QCD,}
  JHEP {07} (2016) 143
  [1602.08105].

\bibitem{threshold}
  S.~Kim and M.~Laine,
  {\it On thermal corrections to near-threshold annihilation,}
  JCAP {01} (2017) 013
  [1609.00474].

\bibitem{idm}
  S.~Biondini and M.~Laine,
  {\it Re-derived overclosure bound for the inert doublet model,}
  JHEP {08} (2017) 047
  [1706.01894].

%
%
\bibitem{giv}
  M.~Garny, A.~Ibarra and S.~Vogl,
  {\it Signatures of Majorana dark matter with $t$-channel mediators,}
  Int.\ J.\ Mod.\ Phys.\ D {24} (2015) 1530019
  [1503.01500].


%
%
\bibitem{hg}
  H.~Goldberg,
  {\it Constraint on the Photino Mass from Cosmology,}
  Phys.\ Rev.\ Lett.\  {50} (1983) 1419;
  {\it ibid.} {103} (2009) 099905 (E). 


%
%
\bibitem{eoz}
  J.~Ellis, K.A.~Olive and J.~Zheng,
  {\it The Extent of the Stop Coannihilation Strip,}
  Eur.\ Phys.\ J.\ C {74} (2014) 2947
  [1404.5571].

%
%
\bibitem{hhk}
  J.~Harz, B.~Herrmann, M.~Klasen,
  K.~Kova\v{r}\'ik and M.~Meinecke,
  {\it SUSY-QCD corrections to stop annihilation into electroweak final
  states including Coulomb enhancement effects,}
  Phys.\ Rev.\ D {91} (2015) 034012
  [1410.8063].

%
%
%
%
\bibitem{ipsv}
  A.~Ibarra, A.~Pierce, N.R.~Shah and S.~Vogl,
  {\it Anatomy of Coannihilation with a Scalar Top Partner,}
  Phys.\ Rev.\ D {91} (2015) 095018
  [1501.03164].



%
%
%
\bibitem{ll}
  S.P.~Liew and F.~Luo,
  {\it Effects of QCD bound states on dark matter relic abundance,}
  JHEP {02} (2017) 091
  [1611.08133].

%
%
%
%
%
\bibitem{mrss}
  A.~Mitridate, M.~Redi, J.~Smirnov and A.~Strumia,
  {\it Cosmological Implications of Dark Matter Bound States,}
  JCAP {05} (2017) 006
  [1702.01141].

%
%
\bibitem{klz}
 W.Y.~Keung, I.~Low and Y.~Zhang,
 {\it A Reappraisal on Dark Matter Co-annihilating with a Top/Bottom
 Partner,}
 Phys.\ Rev.\ D {96} (2017) 015008
 [1703.02977].


\bibitem{dhr}
  J.F.~Donoghue, B.R.~Holstein and R.W.~Robinett,
  {\it Quantum Electrodynamics at Finite Temperature,}
  Annals Phys.\  164 (1985) 233; 
  {\it ibid.} 172 (1986) 483 (E).

\bibitem{ht1}
  R.D.~Pisarski,
  {\it Scattering Amplitudes in Hot Gauge Theories,}
  Phys.\ Rev.\ Lett.\  {63} (1989) 1129.

\bibitem{ht2}
  J.~Frenkel and J.C.~Taylor,
  {\it High Temperature Limit of Thermal QCD,}
  Nucl.\ Phys.\ B {334} (1990) 199.

\bibitem{ht3}
  E.~Braaten and R.D.~Pisarski,
  {\it Soft Amplitudes in Hot Gauge Theories: a General Analysis,}
  Nucl.\ Phys.\ B {337} (1990) 569.

\bibitem{ht4}
  J.C.~Taylor and S.M.H.~Wong,
  {\it The Effective Action of Hard Thermal Loops in QCD,}
  Nucl.\ Phys.\ B {346} (1990) 115.

\bibitem{lsb}
  L.S.~Brown and R.F.~Sawyer,
  {\it Nuclear reaction rates in a plasma,}
  Rev.\ Mod.\ Phys.\  {69} (1997) 411
  [astro-ph/9610256].

\bibitem{imV}
  M.~Laine, O.~Philipsen, P.~Romatschke and M.~Tassler,
  {\it Real-time static potential in hot QCD,}
  JHEP {03} (2007) 054
  [hep-ph/0611300].

\bibitem{bbr}
  A.~Beraudo, J.-P.~Blaizot and C.~Ratti,
  {\it Real and imaginary-time $Q\overline{Q}$
  correlators in a thermal medium,}
  Nucl.\ Phys.\ A {806} (2008) 312
  [0712.4394].

\bibitem{jacopo}
  N.~Brambilla, J.~Ghiglieri, A.~Vairo and P.~Petreczky,
  {\it Static quark-antiquark pairs at finite temperature,}
  Phys.\ Rev.\ D {78} (2008) 014017
  [0804.0993].

\bibitem{resum}
  M.~Laine,
  {\it A Resummed perturbative estimate for the quarkonium spectral 
  function in hot QCD,}
  JHEP {05} (2007) 028
  [0704.1720].

\bibitem{jacopo2}
  N.~Brambilla, M.A.~Escobedo, J.~Ghiglieri and A.~Vairo,
  {\it Thermal width and gluo-dissociation of quarkonium in pNRQCD,}
  JHEP {12} (2011) 116
  [1109.5826].

\bibitem{ghlv}
  M.~Garny, J.~Heisig, B.~L\"ulf and S.~Vogl,
  {\it Coannihilation without chemical equilibrium,}
  Phys.\ Rev.\ D {96} (2017) 103521
  [1705.09292].

\bibitem{bodwin}
 G.T.~Bodwin, E.~Braaten and G.P.~Lepage,
 {\it Rigorous QCD analysis of inclusive annihilation and 
  production of heavy quarkonium,}
  Phys.\ Rev.\ D {51} (1995) 1125; 
  {\it ibid.} {55} (1997) 5853 (E)
  [hep-ph/9407339].

\bibitem{clas1}
  B.W.~Lee and S.~Weinberg, 
  {\it Cosmological Lower Bound on Heavy Neutrino Masses}, 
  Phys.\ Rev.\ Lett.\  {39} (1977) 165.

\bibitem{clas2}
  J.~Bernstein, L.S.~Brown and G.~Feinberg,
  {\it The Cosmological Heavy Neutrino Problem Revisited,}
  Phys.\ Rev.\ D {32} (1985) 3261.

\bibitem{old1}
  K.~Griest and D.~Seckel,
  {\it Three exceptions in the calculation of relic abundances,}
  Phys.\ Rev.\ D {43} (1991) 3191.

\bibitem{chemical}
  D.~B\"odeker and M.~Laine,
  {\it Heavy quark chemical equilibration rate as a transport coefficient,}
  JHEP {07} (2012) 130
  [1205.4987].

\bibitem{Ghi-Sch}
  I.~Ghisoiu, J.~M\"oller and Y.~Schr\"oder,
  {\it Debye screening mass of hot Yang-Mills theory to three-loop order,}
  JHEP {11} (2015) 121
  [1509.08727].

\bibitem{hp}
  J.~Harz and K.~Petraki,
  {\it Higgs Enhancement for the Dark Matter Relic Density,}
  1711.03552.

\bibitem{bkr}
  Y.~Burnier, O.~Kaczmarek and A.~Rothkopf,
  {\it Static quark-antiquark potential in the quark-gluon plasma
  from lattice QCD,}
  Phys.\ Rev.\ Lett.\  {114} (2015) 082001
  [1410.2546].

\bibitem{planck}
  P.A.R.~Ade {\it et al.} [Planck Collaboration],
  {\it Planck 2015 results. XIII. Cosmological parameters,}
  Astron.\ Astrophys.\  {594} (2016) A13
  [1502.01589].

\bibitem{crossover}
  M.~Laine and M.~Meyer,
  {\it Standard Model thermodynamics across the electroweak crossover,}
  JCAP {07} (2015) 035
  [1503.04935].

\bibitem{dono}
  M.~D'Onofrio and K.~Rummukainen,
  {\it Standard model cross-over on the lattice,}
  Phys.\ Rev.\ D {93} (2016) 025003
  [1508.07161].

\bibitem{sf}
  A.~Pierce, N.R.~Shah and S.~Vogl,
  {\it Stop Co-Annihilation in the Minimal Supersymmetric Standard Model
  Revisited,}
  Phys.\ Rev.\ D {97} (2018) 023008
  [1706.01911].

\bibitem{alph1}
  P.A.~Baikov, K.G.~Chetyrkin and J.H.~K\"uhn,
  {\it Five-Loop Running of the QCD coupling constant,}
  Phys.\ Rev.\ Lett.\  {118} (2017) 082002
  [1606.08659].

\bibitem{alph2}
  F.~Herzog, B.~Ruijl, T.~Ueda, J.A.M.~Vermaseren and A.~Vogt,
  {\it The five-loop beta function of Yang-Mills theory with fermions,}
  JHEP {02} (2017) 090
  [1701.01404].

\bibitem{alph3}
  T.~Luthe, A.~Maier, P.~Marquard and Y.~Schr\"oder,
  {\it The five-loop Beta function for a general gauge group and
  anomalous dimensions beyond Feynman gauge,}
  JHEP {10} (2017) 166
  [1709.07718].

\bibitem{pot1}
  Y.~Schr\"oder,
  {\it The Static potential in QCD to two loops,}
  Phys.\ Lett.\ B {447} (1999) 321
  [hep-ph/9812205].

\bibitem{pot4}
  R.N.~Lee, A.V.~Smirnov, V.A.~Smirnov and M.~Steinhauser,
  {\it Analytic three-loop static potential,}
  Phys.\ Rev.\ D {94} (2016) 054029
  [1608.02603].

\bibitem{dr1}
  P. Ginsparg, 
  {\it First and second order phase transitions 
   in gauge theories at finite temperature,}
  Nucl.\ Phys.\ B 170 (1980) 388.

\bibitem{dr2}
  T. Appelquist and R.D. Pisarski,
  {\it High-temperature Yang-Mills theories and three-dimensional 
   Quantum Chromodynamics,}
  Phys.\ Rev.\ D 23 (1981) 2305.

\bibitem{pheneos}
  M.~Laine and Y.~Schr\"oder,
  {\it Quark mass thresholds in QCD thermodynamics,}
  Phys.\ Rev.\ D {73} (2006) 085009
  [hep-ph/0603048].

\bibitem{pdg}
  C.~Patrignani {\it et al.} [Particle Data Group],
  {\it Review of Particle Physics,}
  Chin.\ Phys.\ C {40} (2016) 100001.

\end{thebibliography}
\end{document}